\documentclass[a4paper,10pt]{article}

\usepackage{jheppub} 

\usepackage{bm}
\usepackage{amsmath,amssymb,bm}
\usepackage{graphicx}

\newcommand{\be}{\begin{equation}}
\newcommand{\ee}{\end{equation}}
\newcommand{\bea}{\begin{eqnarray}}
\newcommand{\eea}{\end{eqnarray}}

\newcommand{\eq}[1]{Eq.~(\ref{eq:#1})}
\newcommand{\sect}[1]{Sec.~\ref{sec:#1}}
\newcommand{\appen}[1]{Appendix~\ref{sec:#1}}
\newcommand{\fig}[1]{Fig.~\ref{fig:#1}}

\newcommand{\del}{\partial}

\newcommand{\bra}{\langle}
\newcommand{\ket}{\rangle}
\newcommand{\calO}{{\cal O}}

\newcommand{\eg}{{\it e.g.}}
\newcommand{\ie}{{\it i.e.}}

\newcommand{\bh}{black hole\ }

\newcommand{\SC}{superconductor}
\newcommand{\SCs}{superconductors}
\newcommand{\HSC}{holographic superconductor}
\newcommand{\HSCs}{holographic superconductors}

\bmdefine{\bmq}{{\bm{q}}}
\bmdefine{\bmk}{{\bm{k}}}
\bmdefine{\bmx}{{\bm{x}}}
\bmdefine{\bmy}{{\bm{y}}}
\bmdefine{\bmr}{{\bm{r}}}
\bmdefine{\bmnabla}{{\bm{\nabla}}}
\bmdefine{\bmA}{ \bm{A} }
\bmdefine{\bmD}{ \bm{D} }

\bmdefine{\bmPhi}{ \bm{\Phi} }
\bmdefine{\bmPsi}{ \bm{\Psi} }
\bmdefine{\bmcalO}{ \bm{\mathcal{O}} }
\bmdefine{\bmrho}{ \bm{\rho} }

\newcommand{\calL}{{\cal L}}

\newcommand{\vecx}{\vec{x}}

\newcommand{\epsmu}{\epsilon_{\mu}}

\newcommand{\nj}{\mathfrak{j}}

\newcommand{\Deltap}{\Delta_+}
\newcommand{\Deltam}{\Delta_-}

\bmdefine{\bmg}{{\bm{g}}}
\bmdefine{\bmR}{{\bm{R}}}

\newcommand{\half}{\frac{1}{2}}

\newcommand{\calA}{{\cal A}}

\newcommand{\calF}{{\cal F}}

\newcommand{\calJ}{{\cal J}}
\newcommand{\tilA}{\tilde{A}}

\newcommand{\tilcalA}{\tilde{\calA}}

\newcommand{\tila}{\tilde{a}}
\newcommand{\tilb}{\tilde{b}}

\newcommand{\tilu}{\tilde{u}}
\newcommand{\tilPsi}{\tilde{\Psi}}

\newcommand{\mass}{\text{M}}
\newcommand{\rhou}{U}

\newcommand{\Be}{B_\text{ex}}
\newcommand{\Bc}{B_c}
\newcommand{\Bup}{B_{c2}}
\newcommand{\B}{B}

\newcommand{\increase}{\nearrow}
\newcommand{\decrease}{\searrow}

\newcommand{\Cone}{\delta\psi}

\newcommand{\fGB}{f_\text{\tiny{GB}}}
\newcommand{\la}{\lambda_\text{\tiny{GB}}}
\newcommand{\NGB}{N_\text{\tiny{GB}}}

\newcommand{\Fol}{F_{0\lambda}} 
\newcommand{\Fll}{F_{2\lambda}}
\newcommand{\Atol}{A_t^{(0,\lambda)}}
\newcommand{\Atll}{A_t^{(2,\lambda)}}
\newcommand{\Psiol}{\Psi^{(1,\lambda)}}
\newcommand{\Psill}{\Psi^{(3,\lambda)}}

\newcommand{\muol}{\mu_{0\lambda}}
\newcommand{\mull}{\mu_{2\lambda}}

\newcommand{\ca}{c_a}

\newcommand{\eps}{\epsilon}
\newcommand{\Cnone}{\delta\psi}

\newcommand{\atll}{a_t^{(1,\lambda)}}

\newcommand{\lam}{\lambda_\text{\tiny{coupling}}}

\title{The dual Ginzburg-Landau theory for a holographic superconductor: Finite coupling corrections}

\author[a,1]{Makoto Natsuume%
\note{
Also at Graduate Institute for Advanced Studies, 
SOKENDAI, 1-1 Oho, 
Tsukuba, Ibaraki, 305-0801, Japan;
Department of Mechanical Engineering, Mie University, Tsu, 514-8507, Japan.}
}

\affiliation[a]{KEK Theory Center, \\
Institute of Particle and Nuclear Studies, \\
High Energy Accelerator Research Organization,\\
Tsukuba, Ibaraki, 305-0801, Japan}

\emailAdd{makoto.natsuume@kek.jp}

\arxivnumber{2409.18323}
\preprint{KEK-TH-2609} 

\abstract{
The holographic superconductor is the holographic dual of superconductors. We recently identified the dual Ginzburg-Landau (GL) theory for a class of bulk 5-dimensional holographic superconductors (arXiv:2207.07182 [hep-th]). However, the result is the strong coupling limit or the large-$N_c$ limit. A natural question is how the dual GL theory changes at finite coupling. We identify the dual GL theory for a minimal holographic superconductor at finite coupling (Gauss-Bonnet holographic superconductor), where numerical coefficients are obtained exactly. The GL parameter $\kappa$ increases at finite coupling, namely the system approaches a more Type-II superconductor like material. We also point out two potential problems in previous works: (1) the ``naive" AdS/CFT dictionary, and (2) the condensate determined only from the GL potential terms. As a result,  the condensate increases at finite coupling unlike common folklore. 
}


\begin{document}

\maketitle
\flushbottom

\section{Introduction and Summary}

The AdS/CFT duality or the holographic duality  \cite{Maldacena:1997re,Witten:1998qj,Witten:1998zw,Gubser:1998bc} is a useful tool to study strongly-coupled systems.%
\footnote{See, \eg, Refs.~\cite{CasalderreySolana:2011us,Natsuume:2014sfa,Ammon:2015wua,Zaanen:2015oix,Hartnoll:2016apf,Baggioli:2019rrs} for reviews.}
The \HSC\ is the holographic dual of superconductors \cite{Gubser:2008px,Hartnoll:2008vx,Hartnoll:2008kx},%
\footnote{Strictly speaking, the boundary Maxwell field is not dynamical in the standard \HSC, so one often calls this model the ``holographic superfluid." See remarks below.}
and it may be useful to study strongly-coupled \SCs.

We recently identified the dual GL theory for a class of bulk 5-dimensional \HSCs\ \cite{first}, where numerical coefficients are obtained exactly (see  Refs.~\cite{Herzog:2010vz,Natsuume:2018yrg,Natsuume:2022kic} for related works). However, the result is the strong coupling limit or the large-$N_c$ limit.%
\footnote{Identifying the dual GL theory was initiated in Ref.~\cite{Herzog:2008he} which studied the GL potential terms numerically. Ref.~\cite{Maeda:2009wv} computes all critical exponents and they agree with $|\psi|^4$ mean-field theories. Since then, various works appeared, but they are mostly numerical. See, \eg, Refs.~\cite{Flory:2022uzp,Jeong:2023las} for recent works.}
 A natural question is how the dual GL theory changes at finite coupling. The purpose of this paper is to identify the dual GL theory for a minimal \HSC\ at finite coupling.

The holographic duality has two couplings:
\begin{enumerate}
\item 
The 't~Hooft coupling $\lam$. The $1/\lam$-corrections correspond to higher-derivative corrections or $\alpha'$-corrections. 
\item
The number of ``colors" $N_c$. The $1/N_c$-corrections correspond to string loop corrections or quantum gravity corrections. 
\end{enumerate}
The leading Einstein gravity results are the large-$N_c$ limit, \ie, $\lam\to\infty, N_c\to\infty$. We focus on the former corrections since the latter ones are difficult to evaluate in general and little is known. 

The effect of the $\alpha'$-corrections to the \HSC\ was initiated in Ref.~\cite{Gregory:2009fj}. The previous works show that the condensate takes the value of the mean-field critical exponent. This strongly suggests that the \HSC\ is described by the $|\psi|^4$ mean-field theories even in the presence of the $\alpha'$-corrections. 

However, the exact form of the GL theory is little known.
Also,
\begin{itemize}
\item
Previous works typically compute the condensate and the conductivity and do not compute the other physical quantities (such as the correlation lengths $\xi$, the magnetic penetration length $\lambda$, the critical magnetic fields, and the GL parameter $\kappa$). We would like to know these quantities as well.
\item
In particular, the \HSC\ has the boundary Maxwell field, but in most works, it is not dynamical: one adds it as an external source. This is because one usually imposes the Dirichlet boundary condition on the AdS boundary. As a result, there is no Meissner effect in standard \HSCs. 
Since the Maxwell field is not dynamical, one often calls this case as the ``holographic superfluid."
We impose the ``holographic semiclassical equation" to make the boundary Maxwell field dynamical \cite{Natsuume:2022kic} (see also Ref.~\cite{Compere:2008us}). This makes it possible to discuss the penetration length, the critical magnetic fields, and the GL parameter.
\end{itemize}

We consider a bulk 5-dimensional minimal \HSC\ which corresponds to a 4-dimensional \SC.  As in previous analysis \cite{first}, we consider the bulk scalar mass which saturates the Breitenlohner-Freedman (BF) bound \cite{Breitenlohner:1982bm}. In this case, a simple analytic solution is available at the critical point \cite{Herzog:2010vz}. We compute various physical quantities in the bulk theory and identify the dual GL theory at finite coupling. For example, we evaluate (1) the order parameter response function both at high temperature and at low temperature, (2) the condensate, (3) the  penetration length, (4) the upper critical magnetic field. 

\begin{figure}[tb]
\centering
\scalebox{0.5}{ \includegraphics{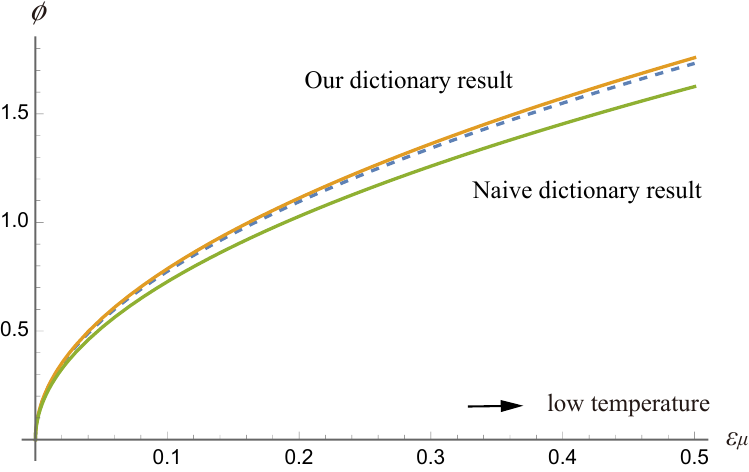} } 

\caption{
The canonically normalized condensate $|\phi|$. The dashed (blue) line represents the strong coupling limit. The solid green and orange lines represent the condensate at finite coupling ($\la=0.1$) using the naive dictionary and our dictionary, respectively.}
\vskip2mm
\label{fig:condensate}
\end{figure}

A \HSC\ is parameterized by a dimensionless parameter $\mu/T$, where $\mu$ is the chemical potential and $T$ is the temperature. We fix $T$ and vary $\mu$.
Our results are summarized by the following GL free energy in the unit $(\pi T)=1$:
\begin{subequations}
\begin{align}
f &= c_0 |D_i\psi|^2-a_0\epsmu|\psi|^2+\frac{b_0}{2}|\psi|^4+\frac{1}{4\mu_m}\calF_{ij}^2-(\psi J^*+\psi^* J)~, \\
D_i &=\del_i-i\calA_i~, \\
c_0 &=\frac{1}{4} \biggl[ 1-\half(25+\pi^2-44\ln2)\la \biggr]~, \\
a_0 &= \half \biggl[ 1-\half(7+\pi^2-12\ln2-12\ln^2 2)\la \biggr]~, \\
b_0 &=\frac{1}{48} \biggl[1-\frac{1}{6}(883+6\pi^2-1176\ln2-216\ln^2 2)\la \biggr]~, \\
\mu_c &= 2+(10-12\ln2)\la~,  \\
\mu_m &=\frac{e^2}{1-c_n e^2}~, \\
c_n&=\left(1-\half\la \right)\ln r_0~.
%
\end{align}
\end{subequations}
Our notations are explained below, but note that this takes the form of the standard GL theory. 
Here,
\begin{itemize}
\item $\epsmu:=\mu-\mu_c$ is the deviation of the chemical potential from the critical point $\mu_c$.
\item The $\la$-terms represent finite coupling corrections. The strong coupling limit is $\la\to0$, and $\lam \propto 1/\la^2$ (\sect{GB}).
\item $e$ is the $U(1)$ coupling, and $\mu_m$ is the magnetic permeability due to the magnetization current or the normal current (\sect{london_alpha'}). The value of $\mu_m$ depends on the boundary condition one imposes.
\item The kinetic term is not canonically normalized which will be important in our analysis.
\item $r_0$ is the horizon radius which is related to $T$ by \eq{temp}. The $T$-dependence is shown explicitly only for the $\ln r_0$ term.
\end{itemize}
Here, we make a few remarks one can extract from the free energy. We assume $\la>0$ (\sect{GB}). 
\begin{itemize}
\item 
The critical point $(\mu/T)_c$ increases at finite coupling. In other words, $T_c$ takes the highest value in the strong coupling limit. 
\item 
A \SC\ is classified by the GL parameter $\kappa$. The GL parameter increases at finite coupling. Namely, the system approaches a Type-II \SC\ like material (\sect{dualGL}).
\item 
 One often says that finite coupling corrections make the condensate ``harder." Namely, the condensate decreases at finite coupling. However, \textit{there are 2 potential problems in previous works and these works must be reexamined} (\sect{naive}): 
\begin{itemize}
\item
First, previous works use the ``naive" AdS/CFT dictionary.
\item
Also, as the above free energy shows, the dual GL theory typically does not have the canonical normalization, and the kinetic term is also corrected if one follows the standard AdS/CFT procedure. Then, \textit{whether the condensate decreases or not should be judged after one normalizes the kinetic term,} \eg, the canonical normalization.  
\end{itemize}
Most previous works do not consider these issues, so their results must be reexamined. If one takes these issues into account, the condensate actually \textit{increases} at finite coupling for our system (\fig{condensate}). This implies that the correlation lengths $\xi$ increase at strong coupling. This is more natural since the correlation between longer distance should be possible at strong coupling.
\end{itemize}

\subsection{Two problems in previous works}\label{sec:naive}

Previous works on finite coupling corrections typically have 2 problems. Here, we point out these problems before we discuss them in details in \sect{dict}. Most previous works do not consider thees issues, so their results must be reexamined. 

First,
\begin{center}
\fbox{
\begin{tabular}{l}
Problem 1: The operator expectation values are extracted \\
by a naive AdS/CFT dictionary.
\end{tabular}
}
\end{center}
%
%
%
In AdS/CFT, one solve the 5-dimensional field equation of a bulk field, and one extracts an operator expectation value form the asymptotic behavior of a bulk field. For example, the asymptotic behavior of our bulk scalar field $\Psi$ is given by
\begin{align}
\Psi \sim \frac{J}{2} z\ln z-\psi\, z+\cdots~, \quad (z\to 0)~,
\label{eq:naive}
\end{align}
where $z:=L^2/r^2$, $r$ is the AdS radial coordinate, and $L$ is the AdS radius. We set $L=1$ for simplicity. One would regard $\psi$ as the condensate and $J$ as the source of the condensate for the dual theory.
 Such a relation is called an AdS/CFT dictionary. 

But this is not a law; rather it is an empirical relation. It is valid  \textit{if the asymptotic behavior of the metric is the same as the pure AdS geometry}. The standard form of the Gauss-Bonnet (GB) \bh does not take the form, so one should reexamine the AdS/CFT dictionary.
This remark also applies if one uses the other \bh backgrounds. 

Usually, the difference is an overall numerical factor and is relatively harmless. But the AdS/CFT dictionary gets finite coupling corrections, so one needs to take them into account. 

This is a well-known fact among experts, but we would like to stress its importance. This completely changes qualitative behaviors for our system. 
We compute physical quantities by our dictionary (\sect{bulk_alpha'}) and by the naive dictionary (\appen{ala_soda}). 
For our system, the qualitative behaviors of most physical quantities become opposite, so the use of the correct dictionary overwhelms the other background effects. 

We derive the AdS/CFT dictionary in \sect{dict}, but it is given by
\begin{subequations}
\label{eq:dict0}
\begin{align}
\Psi &\sim \frac{J}{2} z\ln z - \NGB\psi\, z+\cdots~,\\
\NGB &\sim 1-\half\la~,
\end{align}
\end{subequations}
for our system. 
Note that there is a factor  $\NGB$ which has the $O(\la)$ contribution. This will be important in our discussion.

The other problem is
\begin{center}
\fbox{
\begin{tabular}{l}
Problem 2: The condensate is determined only from the GL potential.
\end{tabular}
}
\end{center}
From the GL point of view, our free energy takes the form
\begin{align}
f &= c |D_i\psi|^2-a|\psi|^2+\frac{b}{2}|\psi|^4~. 
%
\end{align}
Previous works compute the condensate from the potential analysis, so it corresponds to compute
\begin{align}
\epsilon^2 = \frac{a}{b}~.
%
\end{align}
However, the kinetic term of the dual GL theory typically does not have the canonical normalization, \ie, $c\neq1$ if one follows the standard AdS/CFT procedure. In such a case, the``condensate" $\epsilon$ itself is not a physical quantity.
For example, the boundary Maxwell field mass is not $\epsilon^2$ but $c\epsilon^2$. Then, the London penetration length $\lambda$ is given by
\begin{align}
\lambda^2=\frac{1}{2\mu_mc \epsilon^2}~.
%
\end{align}
In such a case, one should regard
\begin{align}
\tilde{\epsilon}^2= \frac{a}{b}c
%
\end{align}
as the real condensate. 

Usually, the difference is an overall numerical factor and is relatively harmless. But $c$ gets a correction at finite coupling, and one needs to take it into account.




\section{Preliminaries}

\subsection{The GB black hole}\label{sec:GB}

The \HSC\ is a gravity-Maxwell-complex scalar system. 
We take the probe limit where the backreaction of the matter fields onto the geometry is ignored.
Then, the matter fields decouple from gravity, so the background solution is a pure gravity solution, and we solve the bulk matter equations in the background.

From string theory point of view, the bulk action is an effective action expanded in the number of derivatives. Schematically,
\begin{align}
S = \int d^{5}x \sqrt{-g} \{ \calL_2+ \calL_4 +\cdots \}~,
\label{eq:bulk_schematic}
\end{align}
where $\calL_i$ denotes $i$-derivative terms. $\calL_2$ is the leading order Lagrangian: for pure gravity, 
\begin{align}
\calL_2=R-2\Lambda~, \quad \Lambda =-\frac{6}{L^2}~,
%
\end{align}
where $L$ is the AdS radius. Then, the background solution is given by the Schwarzschild-AdS$_5$ (SAdS$_5$) black hole:
\begin{subequations}
\begin{align}
ds_5^2 &= \frac{r^2}{L^2}(-fdt^2+d\vecx_3^2)+L^2\frac{dr^2}{r^2f}~, \\
f &= 1-\left(\frac{r_0}{r}\right)^4~.
%
\end{align}
\end{subequations}

We focus on the first nontrivial corrections with four derivatives. In general, one should include all possible independent terms%
\footnote{We use upper-case Latin indices $M, N, \ldots$ for the 5-dimensional bulk spacetime coordinates and use Greek indices $\mu, \nu, \ldots$ for the 4-dimensional boundary coordinates. The boundary coordinates are written as  $x^\mu = (t, x^i) =(t, \vecx)=(t,x,y,z)$. 
}
\begin{align}
%
\calL_4=\half\la L^2(R^2-4R_{AB}R^{AB}+R_{ABCD}R^{ABCD})~,
\label{eq:GB}
\end{align}
where $\la \sim \alpha'/L^2 \ll 1$. 
This particular combination is known as Gauss-Bonnet (GB) combination.%
\footnote{A field redefinition changes $R^2$ and $R_{AB}^2$ but does not change $R_{ABCD}^2$. Thus, only $R_{ABCD}^2$ is the meaningful quantity, but this combination is useful because the Einstein equation is at most second order in derivatives.}
The value $\la$ depends on the theory one considers, but we assume that such a theory exists. 

Note that the convention of the 't~Hooft coupling $\lam$ and $\la$ is a little confusing:
\begin{itemize}
\item
From the boundary point of view, $\lam$ represents the coupling constant of the dual field theory, and $\lam\to\infty$ is the strong coupling limit.
\item
On the other hand, if one uses $\la$, $\la\to0$ is the strong coupling limit. 
\end{itemize}
For the $\mathcal{N}=4$ super-Yang-Mills (SYM) theory, $\lam$ is given by 
\begin{align}
\lam=\left(\frac{L}{l_s}\right)^4~,
%
\end{align}
where $\alpha'=l_s^2$. Thus, $\lam \propto 1/\la^2$, and the strong coupling limit corresponds to $\lam\to \infty$ or $\la\to0$. 

The black hole background of Gauss-Bonnet gravity is obtained in Ref.~\cite{Cai:2001dz}. 
A standard form of Gauss-Bonnet AdS \bh is written as 
\begin{subequations}
\label{eq:GB_bh}
\begin{align}
ds_5^2 &=-F\NGB^2 dt^2+ \left(\frac{r}{L} \right)^2 d\vecx_3^2+\frac{dr^2}{F}~, \\
F&= \frac{r^2}{L^2}\frac{1}{2\la} \left\{1-\sqrt{1-4\la \biggl\{1-\left(\frac{r_0}{r}\right)^4 \biggr\} } \right\}~.
%
\end{align}
\end{subequations}
The constant $\NGB$ may be chosen so that the boundary metric takes the form $ds^2=r^2(-dt^2+d\vecx_3^2)$. 
In the limit $\la\to0$, the metric reduces to the SAdS$_5$ black hole if $\NGB=1$:
\begin{align}
F \to \frac{r^2}{L^2} \biggl\{1-\left(\frac{r_0}{r}\right)^4 \biggr\}~.
%
\end{align}
The thermodynamic quantities are
\begin{subequations}
\begin{align}
\pi T &= \NGB  \frac{r_0}{L^2}~, \\
s &= \frac{1}{4G} \left( \frac{r_0}{L} \right)^3~, \\
\varepsilon &= \NGB \frac{3}{16\pi GL} \left( \frac{r_0}{L} \right)^4~.
%
\end{align}
\end{subequations}

\paragraph{The choice of $\NGB$:}
The $g_{tt}$-component behaves as  
\begin{align}
F\NGB^2 \sim \frac{r^2}{L^2}(1+\la)\NGB^2~, \quad(u\to0)~,
%
\end{align}
so one would choose $\NGB$ so that the boundary metric remains the Minkowski form:
\begin{align}
\NGB^2 \sim 1-\la~.
%
\end{align}
%
%
Then, the Hawking temperature becomes
\begin{align}
\pi T L = \NGB  \frac{r_0}{L}
\sim \left( 1-\half\la \right) \frac{r_0}{L}~.
\label{eq:temp}
\end{align}
Namely, the Hawking temperature gets the $O(\la)$ correction. 
%
%

On the other hand, Ref.~\cite{Gregory:2009fj} sets $\NGB=1$ in \eq{GB_bh} and introduces the ``effective" AdS scale which is $\la$-dependent:
\begin{align}
L_\text{eff}^2\sim \frac{L^2}{1+\la}\sim (\NGB L)^2~.
%
\end{align}
In this case, the Hawking temperature remains the same as the SAdS$_5$ \bh $\pi T'L=r_0/L$. 
The temperature is written as $T'$ to emphasize that $T$ and $T'$ differ.
It is a useful concept to some extent, but we mainly use $\NGB$ and $L$ separately because not all quantities are written in the combination $L_\text{eff}=\NGB L$.

\paragraph{The sign of $\la $:}
Most literature on this system assumes $\la>0$. Following the tradition, we also assume $\la>0$. However, we study the $O(\la)$ corrections, so the qualitative behaviors of the GB \HSC\ depend on the sign of $\la$. Here, we make a few general remarks about the sign of the $\alpha'$ corrections:
\begin{itemize}
\item
For the heterotic string theory, $\la>0$.
\item 
The weak gravity conjecture roughly states that gravity is the weakest force. For an extreme black hole, this implies the relation $M/Q<1$ \cite{Kats:2006xp,Natsuume:1994hd}. For the 5-dimensional Einstein-Maxwell-GB gravity, the entire region of $\la<0$ is excluded from the relation.
\item
However, these are different theories from the system we consider. For example, these are theories on the flat spacetime. The simplest system in AdS/CFT is the $\mathcal{N}=4$ SYM. The gravity dual of the $\mathcal{N}=4$ SYM does not have the $O(\la)$ correction due to supersymmetry. The leading correction is $O(\la^3)$. 
\item
Because GB gravity does not arise for the $\mathcal{N}=4$ SYM, people studied the possible $\la$ from physical consideration. There is a well-known bound from the boundary point of view \cite{Brigante:2008gz,Buchel:2009tt,Hofman:2009ug}. For the 5-dimensional bulk theory, 
\begin{align}
-\frac{7}{36} \leq \la \leq \frac{9}{100}~,
%
\end{align}
so $\la<0$ is not excluded.
\end{itemize}
From the bound, a large $\la$ seems to be excluded. However, the 't~Hooft coupling is $\lam \propto 1/\la^2$, and it should be possible to take the weak coupling limit $\lam\to0$ in principle. In any case,  GB gravity is just the $O(\la)$ corrections. For a large enough $\la$, the $O(\la^2)$ corrections and higher are not negligible. In fact, many of our quantities become problematic when $|\la|\gtrsim O(1)$. 

\paragraph{A caution:}
The solution \eqref{eq:GB_bh} is the exact solution of GB gravity. However, from string theory point of view, one should not take the solution at face value. As indicated in \eq{bulk_schematic}, GB gravity represents just the first-order correction to the bulk action. Thus, one should truncate various results at $O(\la)$. A quick numerical analysis indicates that various results deviate from our exact results if one does not take this point into account. It is unclear to us if previous works take this point into account, which is one motivation of this work. In this paper, we take this point into account and truncate various results at $O(\la)$ consistently. 

\subsection{The holographic superconductor}

We consider the bulk 5-dimensional ``minimal" holographic superconductor:
\begin{subequations}
\begin{align}
%
S_\text{m} &= -\frac{1}{g^2} \int d^5x \sqrt{-g} \biggl\{ \frac{1}{4}F_{MN}^2 + |D_M\Psi|^2+m^2 |\Psi|^2 \biggr\}~, \\
F_{MN} &=\del_M A_N -\del_N A_M~, \quad
D_M =\nabla_M-iA_M~.
%
\end{align}
\end{subequations}
In this paper, we choose the mass dimension as $[A_M]=[\Psi]=\mass$ and $[g^2]=\mass^{-1}$. 
We consider the \HSC\ on the GB \bh background. This kind of \HSC\ is called the Gauss-Bonnet \HSC.
The bulk matter equations are given by
\begin{subequations}
\begin{align}
0 &= D^2\Psi-m^2\Psi~, \\
0 &= \nabla_NF^{MN} - J^M~,\\
J_M &= -i\{ \Psi^* D_M\Psi -\Psi(D_M\Psi)^*\} 
= 2\Im(\Psi^* D_M\Psi)~.
%
\end{align}
\end{subequations}
At high temperature, the bulk matter equations admit a solution
\begin{align}
A_t = \mu(1-u)~, \quad
A_i = 0~, \quad
\Psi =0~,
\label{eq:sol_high}
\end{align}
where $\mu$ is the chemical potential.
A \HSC\ has 2 dimensionful quantities $T$ and $\mu$, so the system is parameterized by a dimensionless parameter $\mu/T$. We fix $T$ and vary $\mu$. The $\Psi=0$ solution becomes unstable at the critical point and is replaced by a $\Psi\neq0$ solution. When the background is the SAdS$_5$ black hole,  there exists a simple analytic solution at the critical point for $m^2=-4/L^2$ \cite{Herzog:2010vz}:
\begin{align}
\Psi \propto -\frac{u}{1+u}~,  \quad\text{at}\quad \left(\frac{\mu}{\pi T}\right)_c=2~.
%
\end{align}
Below we utilize this solution to explore the system.

\paragraph{The asymptotic behavior:}
We rewrite the GB \bh metric  as
\begin{subequations}
\label{eq:GB_u}
\begin{align}
ds^2 &=\left(\frac{\pi T L}{\NGB}\right)^2 \frac{1}{u}(-\fGB\NGB^2 dt^2+d\vecx_3^2)+L^2\frac{du^2}{4u^2\fGB}~, \\
\fGB&= \frac{1}{2\la} \left\{1-\sqrt{1-4\la(1-u^2)} \right\}~,
%
\end{align}
\end{subequations}
where $u=r_0^2/r^2$. 
The asymptotic behavior is 
\begin{align}
ds^2 &\sim \left(\frac{\pi T L}{\NGB}\right)^2 \frac{1}{u}(-dt^2+d\vecx_3^2)+L_\text{eff}^2\frac{du^2}{4u^2}~,
\quad(u\to0)~.
%
\end{align}

For $\Psi=\Psi(u)$, the field equation is given by
\begin{align}
0&= \del_u\biggl( \frac{\fGB}{u}\del_u\Psi \biggr) -\frac{m^2L^2}{4u^3} \Psi~.
%
\end{align}
Asymptotically, 
\begin{align}
0 \sim -\frac{m^2L^2(1-\la)}{4u^2} \Psi -\frac{1}{u}\Psi'+\Psi''~,
%
\end{align}
so the asymptotic behavior is given by
\begin{subequations}
\begin{align}
\Psi &\sim A u^{\Delta_-/2} + B u^{\Delta_+/2}~,
\quad (u\to0)~, \\
\Delta_\pm &= 2\pm\sqrt{4+m^2L^2(1-\la)}~,
%
\end{align}
\end{subequations}
where $(\Delta_+, \Delta_-)$ are the scaling dimensions of the condensate and the source of the condensate, respectively. 
Namely, the $\alpha'$-correction changes the scaling dimension of the condensate $\Delta_+$. Ref.~\cite{Gregory:2009fj} pointed out 2 options: 
\begin{enumerate}
\item One would fix the scaling dimension of the order parameter,
\item Or one would fix the bulk scalar mass $m$.
\end{enumerate}
According to Ref.~\cite{Gregory:2009fj}, they studied both options, and the qualitative behavior does not change, but the reference mainly focuses on the fixed mass. From the boundary point of view, the observable is the scaling dimension, so we prefer to fix the scaling dimension. In fact, the scaling dimension of their condensate changes as one varies $\la$.

So, we keep to fix $\Delta_+=2$:
\begin{align}
4+m^2L^2(1-\la)=0 \to m^2L^2\sim -4(1+\la)~.
%
\end{align}
In this case, the asymptotic behavior is
\begin{align}
\Psi \sim A u \ln u+ B u~, \quad(u\to0)~.
%
\end{align}

\paragraph{Technical differences from Ref.~\cite{Gregory:2009fj}:}
To summarize the technical differences from Ref.~\cite{Gregory:2009fj},
\begin{enumerate}
\item We introduce $\NGB$ so that the boundary metric remains the Minkowski form $ds^2=-dt^2+d\vecx^2$.
\item We use a different mass $m^2$ for the bulk scalar field $\Psi$.
\item We fix the scaling dimension of the order parameter instead of fixing the bulk scalar mass. 
\item We express our results using $\mu$ instead of using the charge density $\rho$. 
\item We take into account the modification of the AdS/CFT dictionary in the GB \bh background (\sect{dict}).
\end{enumerate}

\paragraph{Counterterms:}
In the bulk 5-dimensions, one needs the counterterm action for the Maxwell field to cancel the UV divergences. In the presence of the $\la$-corrections, we choose the counterterm as
\begin{align}
S_\text{CT}= -\int d^4x\, \frac{1}{4g^2}\sqrt{-\gamma}\gamma^{\mu\nu}\gamma^{\rho\sigma}F_{\mu\rho}F_{\nu\sigma} \times (\NGB L)\ln\left(\frac{L}{r_0}u^{1/2}\right)~,
\label{eq:CT}
\end{align}
where $\gamma_{\mu\nu}$ is the 4-dimensional boundary metric (the 4-dimensional part of the bulk metric). 
%
%
Note that the counterterm has the $O(\la)$ correction. Also, the log term takes the form $\ln\tilu$ if one uses $\tilu:=L/r$. We use $u=(r_0/r)^2=(r_0/L)^2 \tilu^2$, so $\ln \tilu=\ln(u^{1/2}L/r_0)$. 
%
%

\section{Qualitative analysis}\label{sec:formal}

\subsection{The critical point}\label{sec:critical}

Below we set $(\pi T)=1, g=L=1$. First, consider the critical point and its  solution. We approach the critical point from the high-temperature phase.
At high temperatures, the background solution is given by \eq{sol_high}.
Consider the linear perturbation from the background $\Psi= 0+\delta\Psi$. 
When $\Psi=0$, $\delta A_t$ and $\delta A_i$ decouple from the $\delta\Psi$-equation, and it is enough to consider the $\delta\Psi$-equation:
\begin{align}
0&= \del_u\biggl( \frac{\fGB}{u}\del_u\delta\Psi \biggr) + \biggl[ \frac{A_t^2}{4 \fGB u^2} 
- \frac{m^2}{4u^3} \biggr] \delta\Psi~.
%
\end{align}
%
%

We construct the homogeneous solution at the critical point by the $\la$-expansion:
\begin{subequations}
\begin{align}
A_t &= \mu_0(1-u)~, \\
\delta\Psi &= F_0+ \la \Fol+\cdots~. 
%
\end{align}
\end{subequations}
When $\la=0$, the spontaneous condensate solution is
\begin{align}
F_0= - \NGB^3\Cone\,\frac{u}{1+u} \sim -\NGB^3\Cone\, u~, \quad (u\to0)~,
%
\end{align}
where $\mu_0=2$. 
As we discuss in \sect{dict}, the AdS/CFT dictionary in the $u$-coordinate is given by
\begin{align}
\delta\Psi \sim \NGB^2\frac{J}{2} u\ln u - \NGB^3\Cone\, u+\cdots~,
%
\end{align}
so $\Cone$ is the condensate. 
But this is the solution when $\la=0$.

At $O(\la)$, $\Psi$ is no longer a spontaneous condensate solution at $\mu_0=2$ and is a solution with the order parameter source. However, it is possible to obtain a source-free solution by choosing $\mu_c$ appropriately. Namely, the effect of $\la$ shifts the critical point as 
\begin{align}
\mu_c=2 \to \mu_c=2+\la\muol+\cdots~.
%
\end{align}

The solution $\Fol$ is given by
\begin{subequations}
\begin{align}
\Fol &= \Cnone\, \frac{u}{2(1+u)} 
 \\
&\times\left[ \pi^2-6 \ln^2 2+10 \ln (1+u)-12 \ln (1-u) \ln\left(\frac{1+u}{2}\right) 
 -6 \text{Li}_2(-u)-12 \text{Li}_2\left(\frac{1+u}{2}\right)\right] 
\nonumber  \\
& \sim O(u^2)~,
\quad(u\to0)~,
\nonumber \\
\muol &=10-12\ln2~.
%
\end{align}
\end{subequations}
Then,
\begin{align}
\left( \frac{\mu}{\pi T} \right)_c = 2+(10-12\ln2)\la \approx 2+1.682\la~, \increase
%
\end{align}
%
%
Here, the arrow indicates the behavior at finite coupling. 
$\mu_c$ becomes larger as one increases $\la$. Namely, $\mu_c$ becomes the minimum or $T_c$ becomes the maximum when $\la=0$. 

\subsection{High-temperature phase}\label{sec:high_formal}

\paragraph{The order parameter response function:}
In the high-temperature phase, there does not exist a spontaneous condensate solution, but there exists a solution with the order parameter source. We consider such a solution here. Namely, we consider the response to the order parameter source and obtain the ``order parameter response function." This gives interesting physical quantities such as the correlation length and the thermodynamic susceptibility.

We consider the perturbation of the form $e^{iqx}$.
The field equation is given by
\begin{align}
0&= \del_u\biggl( \frac{\fGB}{u}\del_u\delta\Psi \biggr) + \biggl[ \frac{A_t^2}{4\fGB u^2} - \NGB^2\frac{q^2}{4u^2} - \frac{m^2}{4u^3} \biggr] \delta\Psi~,
\label{eq:scalar_high1}
\end{align}
where $A_t =(\mu_c+\epsmu)(1-u)$. In the high-temperature phase, $\epsmu<0$. 
Set $\epsmu\to l^2 \epsmu, q\to l q$, and we make a double-series expansion in $l$ and $\la$:
\begin{align}
\delta\Psi &= ( F_0+ \la \Fol+\cdots )+
 l^2 ( F_2+\la \Fll+\cdots )+\cdots~. 
%
\end{align}
We impose the boundary conditions (1) regular at the horizon (2) no fast fall off other than $F_0$.
The latter means that the condensate comes only from $F_0$.
The solutions $F_0$ and $\Fol$ are obtained in \sect{critical}.

The $O(\la^0)$ solution is obtained in Refs.~\cite{Natsuume:2018yrg,Natsuume:2022kic}:
\begin{align}
F_2 & \sim \Cnone\, \frac{q^2-2\epsmu}{8}u \ln u~, \quad(u\to0)~.
%
\end{align}
$\Fll$ would take the form
\begin{align}
%
\Fll \sim \Cnone\, \{ O(q^2)+O(\epsmu)\} u\ln u~.
%
\end{align}
We rerwite the result as
\begin{align}
J &= \frac{1}{4}( c_q q^2 - 2c_a\epsmu ) \Cnone~.
\label{eq:high_formal}
\end{align}
One then obtains the order parameter response function at high temperature $\chi_>$:
\begin{align}
\chi_>=\frac{\del \delta\psi}{\del J} =\frac{ 4}{c_q (q^2-\frac{2c_a}{c_q}\epsmu) }  \propto \frac{1}{q^2+\xi_>^{-2}}~,
%
\end{align}
and the correlation length $\xi_>$, the thermodynamic susceptibility $\chi_>^T$, and the critical amplitude $A_>$ are given by
\begin{subequations}
\begin{align}
\xi_>^2 &:=-q^{-2} = \frac{1}{-2\epsmu}\frac{c_q}{c_a}~, \\
\chi_>^T&:=\chi_>|_{q=0} = \frac{2}{-c_a} =: \frac{A_>}{-\epsmu}~, \\
A_> &=  \frac{2}{c_a}~.
%
\end{align}
\end{subequations}

%
%

\paragraph{The upper critical magnetic field:}
We consider the solution of the form $\Psi=\Psi(\vecx,u), A_t=A_t(\vecx,u), A_y=A_y(\vecx,u)$.  The static bulk equations are given by
\begin{subequations}
\label{eq:bulk_eom}
\begin{align}
0&= \del_u\biggl( \frac{\fGB}{u}\del_u\Psi \biggr) 
+ \biggl[ \frac{A_t^2}{4 \fGB u^2}+ \frac{\NGB^2}{4u^2}(\del_i-iA_i)^2 - \frac{m^2}{4u^3} \biggr] \Psi~,\\
0&= \del_u^2 A_t -\frac{1}{2\fGB u^2}|\Psi|^2 A_t + \frac{\NGB^2}{4\fGB u}\del_i^2 A_t~, \\
0 &=\del_u(\fGB \del_u A_y)+\frac{\NGB^2}{4u}\del_i^2A_y -\frac{|\Psi|^2}{2u^2}A_y + \frac{1}{2u^2} \Im[\Psi^* \del_y\Psi]~,
%
\end{align}
\end{subequations}
where we take the gauge $A_u=0$ and $\del_iA^i=0$. In this gauge, one can set $\Psi=\Psi^*$. 
The index $i$ is raised and lowered by $\delta_{ij}$ not $g_{ij}$.

We apply a magnetic field $B$ and approach the critical point from the high-temperature phase. The scalar field $\Psi$ should have an inhomogeneous condensate at $\Bup$. 

The problem has been discussed in Ref.~\cite{Natsuume:2022kic} in the strong coupling limit. The problem is solved as the Landau problem, and it was pointed out that the problem reduces to the one for the order parameter response function at high temperature with the replacement $q^2\to\Bup$. Thus, \textit{the following relation holds exactly}:
\begin{align}
\Bup = \frac{1}{-\xi_>^2}~.
%
\end{align}
Also, we consider the \HSC\ with a particular scalar mass $m$, but \textit{the above relation holds exactly for the minimal \HSC\ with arbitrary mass.} 

Even at finite coupling, the situation is the same. 
The scalar field equation takes the same form as the strong coupling limit one with the replacement $f\to \fGB$.
Thus, the same argument applies as we repeat below. 

Near $\Bup$, $\Psi$ remains small, and one can expand matter fields as a series in $\epsilon$:
\begin{subequations}
\begin{align}
\Psi(\vecx,u) &= \epsilon\Psi^{(1)}+\cdots~, \\
A_t(\vecx,u) &= A_t^{(0)}+\epsilon^2 A_t^{(2)}+\cdots~, \\
A_y(\vecx,u) &= A_y^{(0)}+\epsilon^2 A_y^{(2)}+\cdots~.
%
\end{align}
\end{subequations}
At  zeroth order,
\begin{align}
A_t^{(0)} = \mu_c(1-u)~,\quad
A_x^{(0)} = 0~,\quad
A_y^{(0)} = \B x~.
%
\end{align}
At  first order, using separation of variables $\Psi^{(1)}=\chi(x)U(u)$, one obtains 
\begin{subequations}
\begin{align}
& (-\del_x^2+B^2x^2)\chi = E \chi~, \\
& \del_u\left( \frac{\fGB}{u}\del_u U \right) + \left[ \frac{(A_t^{(0)})^2}{4\fGB u^2} - \frac{m^2}{4u^3} \right] U = \frac{E\NGB^2}{4u^2}U~,
%
\end{align}
\end{subequations}
where $E$ is a separation constant. The regular solution of $\chi$ is given by Hermite function $H_n$ as
\begin{align}
\chi=e^{-z^2/2}H_n(z)~, z:= \sqrt{B}x~,
%
\end{align}
with the eigenvalue $E=(2n+1)B$. 
$\B$ takes the maximum value when $n=0$ which gives the upper critical magnetic field $B_{c2}$.
Then, the $U$-equation becomes
\begin{align}
0 &= \del_u\left( \frac{\fGB}{u}\del_u \rhou \right) + \left[ \frac{(A_t^{(0)})^2}{4\fGB u^2} - \NGB^2\frac{B_{c2}}{4u^2} - \frac{m^2}{4u^3} \right] \rhou~.
%
\end{align}
To obtain the upper critical magnetic field $\Bup$, we need the source-free solution (spontaneous condensate) for $U$. But the equation is just \eq{scalar_high1} with the replacement $B_{c2}\to q^2$.
Thus, the relation
\begin{align}
\Bup=\frac{1}{-\xi_>^2}
\label{eq:Bc2}
\end{align}
remains valid even at finite coupling. Of course, both quantities should receive $\alpha'$-corrections.


\subsection{Low-temperature background}\label{sec:background_formal}

The solution in \sect{critical} is the one only at the critical point, and we need the background solution in the low-temperature phase.

Consider the solution of the form
\begin{align}
\Psi=\Psi(u)~, \quad A_t=A_t(u)~, \quad A_i=A_u=0~.
%
\end{align}
The field equations are given by
\begin{subequations}
\begin{align}
0&= \del_u\biggl( \frac{\fGB}{u}\del_u\Psi \biggr) + \biggl[ \frac{A_t^2}{4 \fGB u^2}-\frac{m^2}{4u^3} \biggr] \Psi~,\\
0&= \del_u^2 A_t -\frac{1}{2\fGB u^2}|\Psi|^2 A_t~, \\
0&=\Psi^* \Psi'-\Psi^{*'} \Psi~.
%
\end{align}
\end{subequations}
One can set $\Psi$ to be real. 
We make a double-series expansion both in $\eps$ and $\la$:
\begin{subequations}
\begin{align}
A_t(u) &= (A_t^{(0)}+\la \Atol+\cdots)+\eps^2 (A_t^{(2)}+\la \Atll+\cdots)+ \cdots~,\\
\Psi(u) &= \eps( \Psi^{(1)}+ \la\Psiol+\cdots )+
 \eps^3( \Psi^{(3)}+\la\Psill+\cdots )+\cdots~. 
%
\end{align}
\end{subequations}
We impose the following boundary conditions:
\begin{enumerate}
\item $\Psi$: no fast falloff other than $\Psi^{(1)}$. This means that the condensate comes only from $\Psi^{(1)}$. At the horizon, we impose the regularity condition. 
\item $A_t$: $A_t=0$ at the horizon.
\end{enumerate}
Namely, we fix the fast falloff $\psi$, and the chemical potential is corrected:
\begin{align}
A_t^{(0)} \sim \mu_0=2~, \quad
\Atol \sim \muol~, \quad
A_t^{(2)} \sim \mu_2~, \quad
\Atll \sim \mull~.
%
\end{align}
Then, the chemical potential becomes
\begin{subequations}
\begin{align}
\mu&=A_t|_{u=0} \\
&= (2+\la \muol+\cdots)+\eps^2(\mu_2+\la \mull+\cdots)+\cdots~.
%
\end{align}
\end{subequations}
This fixes the overall constant $\epsilon$ of the condensate as
\begin{subequations}
\begin{align}
\epsmu &:= \mu-\mu_c = \eps^2(\mu_2+\la\mull+\cdots)+\cdots~, \\
\to 
\eps^2 &= \frac{\epsmu}{\mu_2+\la \mu_{2\lambda}} \\
&= \frac{1}{\mu_2} c_e \epsmu~,
\label{eq:condensate_formal} \\
c_e &=1-\la\frac{\mu_{2\lambda}}{\mu_2}+\cdots~.
%
\end{align}
\end{subequations}

The condensate satisfies $\epsilon\propto \epsmu^{1/2}$, so the critical exponent remains the same as the strong coupling limit and takes the mean-field value $\beta=1/2$. This is a well-known result, but this is clear in this analysis. This is due to the large-$N_c$ limit. In order to obtain a non-mean-field value, one needs to take into account $1/N_c$-corrections not $\alpha'$-corrections.


\section{The dual GL theory}\label{sec:dualGL}

We consider the following GL theory:
\begin{subequations}
\label{eq:GL4}
\begin{align}
f &= c_K|D_i\psi|^2 - a |\psi|^2 + \frac{b}{2}|\psi|^4+\cdots+\frac{1}{4\mu_m} \calF_{ij}^2  -(\psi J^*+\psi^* J)~,  \\
D_i &= \del_i - i\calA_i~, \quad
a = a_0\epsmu+\cdots~, \quad
b =b_0+\cdots~, \quad
c_K =c_0+\cdots~.
\end{align}
\end{subequations}
In the standard GL theory, $\mu_m=e^2$. Namely, we generalize the GL theory where the material has the magnetization current.
The equations of motion are given by
\begin{subequations}
\begin{align}
0&=-c_KD^2\psi-a\psi+b\psi|\psi|^2-J~,\\
0&=\del_j \calF^{ij}-\mu_m \calJ^i~,\\
\calJ_i &=-ic_K[\psi^* D_i\psi - \psi (D_i\psi^*)] = 2c_K\Im[\psi^* D_i\psi]~.
%
\end{align}
\end{subequations}
There are 3 unknown coefficients $a_0,b_0,c_0$.
The coefficient $c_0$ is actually redundant because one can always absorb it by a $\psi$ scaling. Thus, there are 2 independent parameters. But it is useful to keep it to compare with holographic results. It turns out that the dual GL theory does not have the canonical normalization if one uses the standard AdS/CFT dictionary. Namely, the scaling changes the AdS/CFT dictionary. 
We take the scaling into account after we obtain the final result (\sect{canonical}).

We determine these coefficients of the dual GL theory from (1) the order parameter response function at high temperature, and (2) the spontaneous condensate.

In the high-temperature phase $\epsmu<0$, there is no spontaneous condensate. When there is no Maxwell field, the linearized $\psi$ equation is 
\begin{align}
J=-c_0 \del_i^2 \psi-a_0\epsmu\psi~.
%
\end{align}
In the momentum space where $\psi\propto e^{iqx}$, 
\begin{align}
J=(c_0 q^2-a_0\epsmu)\psi
\stackrel{\text{bulk}}{=} \frac{1}{4}(c_q q^2 -2 c_a \epsmu) \psi~,
%
\end{align}
where the last expression is the formal bulk result \eqref{eq:high_formal}.
Thus,
\begin{align}
c_0 =\frac{1}{4}c_q~, \quad
a_0 =\half \ca~.
%
\end{align}
In the low-temperature phase $\epsmu>0$, there exists a homogeneous spontaneous condensate:
\begin{align}
|\psi_0|^2 = \epsilon^2=\frac{a_0}{b_0}\epsmu 
\stackrel{\text{bulk}}{=} \frac{1}{\mu_2}c_e\epsmu~, 
%
\end{align}
where the last expression is the formal bulk result \eqref{eq:condensate_formal}.
Thus,
\begin{align}
b_0 = \frac{a_0}{c_e}\mu_2 = \frac{c_a}{2c_e}\mu_2~.
%
\end{align}
Here, $\mu_2=1/24$ \cite{Herzog:2010vz,Natsuume:2018yrg,Natsuume:2022kic}.

In the unit $(\pi T)=1$, the explicit holographic results in \sect{bulk_alpha'} are
\begin{subequations}
\begin{align}
c_q &= 1- \half(25+\pi^2-44\ln2)\la \approx 1-2.186\la~, \\
\ca &= 1- \half (7+\pi^2-12\ln2-12\ln^2 2 )\la \approx 1-1.393\la~,\\
c_e &= 1+ \frac{1}{6}(862+3\pi^2-1140\ln2-180\ln^2 2)\la \approx 1+2.490\la~, \\
\mu_c &= 2+(10-12\ln2)\la \approx 2+1.682\la~, \\
\mu_m &=\frac{e^2}{1-c_n e^2}~, \\
c_n&=\left(1-\half\la\right)\ln r_0 = \left(1-\half\la\right)\ln \left( \frac{\pi T}{\NGB}\right)~. 
%
\end{align}
\end{subequations}
Thus,
\begin{subequations}
\begin{align}
c_0 &=\frac{1}{4} \biggl[ 1-\half(25+\pi^2-44\ln2)\la \biggr]\\
&\approx \frac{1}{4}(1-2.186\la)~, \decrease  \\
a_0 &= \half \biggl[ 1-\half(7+\pi^2-12\ln2-12\ln^2 2)\la \biggr]\\
&\approx \half(1-1.393\la)~, \decrease  \\
b_0 &=\frac{1}{48} \biggl[1-\frac{1}{6}(883+6\pi^2-1176\ln2-216\ln^2 2)\la \biggr]\\
&\approx \frac{1}{48}(1-3.883\la)~. \decrease 
%
\end{align}
\end{subequations}
Here, the arrows indicate the behaviors at finite coupling. 
From the GL theory, one expects the following results. The order parameter response function at high temperature is given by
\begin{subequations}
\begin{align}
\chi_> &=\frac{\del\psi}{\del J} = \frac{1}{c_0q^2-a_0\epsmu} \propto \frac{1}{q^2+\xi_>^{-2}}~, \\
\xi_>^2 &=\frac{c_0}{a_0}\frac{1}{-\epsmu}  \\
& \frac{1}{-2\epsmu}\left[ 1- (9-16\ln2+6\ln^2 2) \la \right] \\
&\approx \frac{1}{-2\epsmu}(1-0.7924\la)~, 
\decrease \\
\chi_>^T &:= \chi_> |_{q=0} = \frac{1}{-a_0 \epsmu} := \frac{A_>}{-\epsmu}~, \\
A_> &= \frac{1}{a_0}~.
%
\end{align}
\end{subequations}
Namely, the correlation length decreases at finite coupling.  This is natural since one expects that the correlation between longer distance is possible at strong coupling.

Then, the upper critical magnetic field is given by
\begin{subequations}
\begin{align}
\Bup &= \frac{a_0}{c_0\epsmu} = \frac{1}{-\xi_>^2} \\
&=2\epsmu\left[ 1+(9-16\ln2+6\ln^2 2) \la \right] \\
&\approx 2\epsmu(1+0.7924\la)~,  
\increase
%
\end{align}
\end{subequations}
using \eq{Bc2}. 
The ``condensate" is given by
\begin{subequations}
\begin{align}
|\psi_0|^2 &= 24\epsmu \left[1+ \frac{1}{6}(862+3\pi^2-1140\ln2-180\ln^2 2)\la \right] \\
&\approx 24\epsmu(1+2.490\la)~. \increase
%
\end{align}
\end{subequations}
At finite coupling, the ``condensate" increases unlike common folklore. The order parameter response function at low temperature is given by
\begin{subequations}
\begin{align}
\chi_< &=\frac{\del\psi}{\del J} = \frac{1}{c_0q^2+2a_0\epsmu} \propto \frac{1}{q^2+\xi_<^{-2}}~, \\
\xi_<^2 &=\frac{c_0}{2a_0\epsmu} 
= \half |\xi_>|^2~, \\
\chi_<^T &:= \chi_< |_{q=0} = \frac{1}{2a_0 \epsmu} := \frac{A_<}{\epsmu}~, \\
A_< &= \frac{1}{2a_0}
= \half A_>~.
%
\end{align}
\end{subequations}
The magnetic penetration length is given by
\begin{align}
\lambda^2 &=\frac{1}{2c_0\mu_m\epsilon^2} = \frac{2}{\mu_m\epsilon^2}c_q~.
%
\end{align}
In terms of $\epsmu$,
\begin{subequations}
\begin{align}
\lambda^2 &=\frac{1}{2c_0\mu_m} \frac{b_0}{a_0\epsmu} \\
&= \frac{1}{12\mu_m\epsmu}\left[ 1-\frac{1}{6}(787-1008\ln2-180\ln^2 2) \la \right] \\
&\approx \frac{1}{12\mu_m\epsmu}(1-0.3043\la)~. 
\decrease 
%
\end{align}
\end{subequations}
Here, we consider a fixed value of $\mu_m$ for simplicity.

A \SC\ has 2 characteristic length scales:
\begin{itemize}
\item The correlation length $\xi_>^2$.
\item The magnetic penetration length $\lambda^2$.
\end{itemize}
Then, a \SC\ is classified by a dimensionless parameter, the GL parameter $\kappa$:
\begin{align}
\kappa^2 &= \frac{\lambda^2}{-\xi_>^2} =\half \left( \frac{B_{c2}}{B_c} \right)^2~.
%
\end{align}
When $\kappa^2<1/2$, the material belongs to a Type I \SC.
When $\kappa^2>1/2$, the material belongs to a Type II \SC.
\begin{subequations}
\begin{align}
\kappa^2 &= \frac{\lambda^2}{-\xi_>^2} = \frac{1}{2\mu_m}\frac{b_0}{c_0^2} \\
&= \frac{1}{6\mu_m}\biggl[ 1+\frac{1}{6}(-733+912\ln2+216\ln^2 2) \la  \biggr] \\
&\approx \frac{1}{6\mu_m}(1+0.4880\la)~, \increase
%
\end{align}
\end{subequations}
At finite coupling, the system approaches a Type-II superconductor like material.

We confirm these results explicitly in next section.
A quantity we are not able to obtain is the thermodynamic critical magnetic field $\Bc$, but the GL prediction is
\begin{subequations}
\begin{align}
\Bc^2 &= \frac{a_0^2}{b_0}\mu_m \\
&= 12\mu_m\epsmu^2\left[ 1+\frac{1}{6}(841-1104\ln2-144\ln^2 2) \la \right] \\ 
&\approx 12\mu_m\epsmu^2(1+1.097\la)~. 
\increase 
%
\end{align}
\end{subequations}

\subsection{The canonical form}\label{sec:canonical}

Our GL theory 
\begin{align}
f &= c_0  |D_i\psi|^2 - a|\psi|^2
+ \frac{b_0}{2}  |\psi|^4 
-(\psi J ^*+\psi^* J) +\cdots
%
\end{align}
does not take the canonical normalization, so rewrite it in the canonical form:
%
%
\begin{subequations}
\begin{align}
|\phi|^2 &= c_0|\psi|^2~, \quad j^2= \frac{1}{c_0} J^2~, \\
f &= |D_i\phi|^2
-\tila|\phi|^2+\frac{\tilb}{2}|\phi|^4
-(\phi j^* +\phi^* j) 
+\cdots~.
%
\end{align}
\end{subequations}
Here,
\begin{subequations}
\begin{align}
\tila = \frac{a}{c_0}
&= 2\epsmu \left[ 1+ (9-16\ln2+6\ln^2 2) \la  \right] \\
&\approx 2\epsmu(1+0.7924\la)~,
\increase \\
\tilb_0 = \frac{b_0}{c_0^2} 
&= \frac{1}{3}\left[ 1+ \frac{1}{6}(-733+912\ln2+216\ln^2 2) \la  \right] \\
&\approx \frac{1}{3}(1+0.488\la)~, \increase 
%
\end{align}
\end{subequations}

Under the scaling, various physical quantities remain the same and behave as
\begin{align}
\xi_>^2 &= \frac{1}{-\tila} 
\decrease~,
\lambda^2 =\frac{\tilb_0}{ 2\mu_m \tila}
\decrease~,
\kappa^2 =\frac{\tilb_0}{2\mu_m} 
\increase~, 
\Bup = \tila 
\increase~,
\Bc^2 = \mu_m\frac{\tila^2}{\tilb_0} 
\increase ~.
\end{align}
but the condensate changes as
\begin{subequations}
\begin{align}
\tilde{\epsilon}^2 =\frac{\tila}{\tilb_0} =\frac{a}{b_0}c_0
&= 6 \epsmu \left[ 1+\frac{1}{6}(787-1008\ln2-180\ln^2 2) \la \right] \\ 
&\approx 6\epsmu(1+0.3043\la)~,
\increase 
%
\end{align}
\end{subequations}
At finite coupling,  $\tila_0$ and $\tilb_0$ increase. This has the following implications:
\begin{enumerate}
\item The correlation lengths $\xi  \propto 1/\tila_0$ decrease at finite coupling.
\item $\Bup\propto \tila_0$ increases.
\item The GL parameter $\kappa^2\propto \tilb_0$ increases. Namely, the system approaches a more Type-II superconductor like material.
\item The GL parameter is also expressed as $\kappa^2=(B_{c2}/B_c)^2/2$, so the ratio $\Bup/\Bc$ increases.
\item The condensate $\epsilon$ depends both on $\tila_0$ and $\tilb_0$ but it increases. One often says that $\la$ makes the condensate ``harder", \ie, $\epsilon$ decreases at finite coupling. But $\epsilon$ actually increases both before and after the canonical normalization.
\item The penetration length $\lambda$ decreases since $\lambda^2\propto 1/\epsilon^2$.
\item $\Bc$ also depends both on $\tila_0$ and $\tilb_0$, but it increases.
\end{enumerate}

\section{Bulk analysis}\label{sec:bulk_alpha'}

\subsection{The AdS/CFT dictionary in the GB background}\label{sec:dict}

In this subsection, we restore dimensionful quantities $L,T$, and $g$ . Previous works on the GB \bh often have problems in the AdS/CFT dictionary. One extracts operator expectation values from the asymptotic behaviors of matter fields, but one uses the naive dictionary such as \eq{naive}. But this is valid  \textit{if the asymptotic behavior of the metric is the same as the pure AdS geometry}. However, the standard form of the GB \bh does not take the form, so the AdS/CFT dictionary should be derived more carefully.
Here, we derive how the AdS/CFT dictionary is modified for the GB black hole, and one needs to take it into account when one discusses finite-coupling corrections. There are several ways to derive the dictionary:
\begin{enumerate}
\item
Here, we simply rewrite the GB \bh so that the asymptotic form coincides with the pure AdS geometry.
\item
Instead, one can directly evaluate the covariant formula such as \eq{current_covariant} in the GB background.
\item
It is always safe to derive the dictionary from fundamental relations such as the GKP-Witten relation \cite{Witten:1998qj,Gubser:1998bc}. Namely, derive the boundary action from the on-shell bulk action.
\end{enumerate}

First, consider the pure AdS geometry:
\begin{subequations}
\begin{align}
ds^2 &= \frac{r^2}{L^2}(-dt^2+d\vecx_3^2)+L^2\frac{dr^2}{r^2} \\
&= \frac{1}{z}(-dt^2+d\vecx_3^2)+L^2\frac{dz^2}{4z^2}~,
%
\end{align}
\end{subequations}
where $z:=(L/r)^2$. The asymptotic behaviors of matter fields are given by%
\footnote{The factor $1/2$ for the slow falloff of the scalar field comes from the fact that we use the coordinate $z\propto r^{-2}$.}
\begin{subequations}
\begin{align}
A_\mu & \sim \tilcalA_\mu  + \tilA_\mu^{(+)} z~, \\
\Psi &\sim \frac{1}{2}\tilPsi^{(-)} z \ln z +\tilPsi^{(+)} z~,
\end{align}
\end{subequations}
where $\tilcalA_t=\mu$ and $\tilcalA_i$ are the chemical potential and the vector potential, respectively.
In this paper, we choose the mass dimensions as $[A_M]=[\Psi]=\mass$ and $[g^2]=\mass^{-1}$, so
\begin{align}
[\tilcalA_\mu]= [\tilcalA_\mu^{(+)}]=[\tilPsi^{(-)}]=[\tilPsi^{(+)}]=\mass~.
%
\end{align}
Using the standard procedure, one obtains
\begin{subequations}
\label{eq:dict_z}
\begin{align}
\bra \calJ^i \ket &= \left. \frac{1}{g^2}\sqrt{-g}F^{zi} \right|_{z=0} +(\text{counterterm}) 
\label{eq:current_covariant} \\
&= \frac{2}{g^2L} \tilA_i^{(+)}+(\text{counterterm})~, \\
\psi =\bra \calO \ket &= -\frac{1}{g^2L} \tilPsi^{(+)}~, \\
J&=\tilPsi^{(-)}~.
%
\end{align}
\end{subequations}
where $\calJ^i$ is the current density, $\psi$ is the order parameter, and $J$ is its source. 
The mass dimension is $[\calJ^i]=\mass^3$ as expected.

Now, consider the GB black hole. The asymptotic behavior is given by
\begin{subequations}
\begin{align}
ds^2 &\sim \left(\frac{\pi T L}{\NGB}\right)^2 \frac{1}{u}(-dt^2+d\vecx_3^2)+L_\text{eff}^2\frac{du^2}{4u^2} \\
&=\frac{1}{z}(-dt^2+d\vecx_3^2)+L_\text{eff}^2\frac{dz^2}{4z^2}~,\\
u &=  \left(\frac{\pi T L}{\NGB}\right)^2 z~.
%
\end{align}
\end{subequations}
Here, $L_\text{eff}:=\NGB L$ is the ``effective" AdS scale. 
%
%
Then, the metric takes the form of the pure AdS with the replacement $L\to L_\text{eff}$. This means that one has to replace the dictionary \eqref{eq:dict_z}  with $L_\text{eff}$ which has an $O(\la)$ correction: 
\begin{subequations}
\label{eq:dictionary_radial}
\begin{align}
\bra \calJ^i \ket &= \frac{2}{g^2\NGB L} \tilA_i^{(+)}+(\text{counterterm})~, \quad (z\text{-coordinate})\\
\psi &= -\frac{1}{g^2\NGB L} \tilPsi^{(+)}~.
%
\end{align}
\end{subequations}
Ref.~\cite{Gregory:2009fj} does not include these corrections.

One would compute various quantities in the $u$-coordinate, transform the results in the $z$-coordinate, and use the above dictionary. Instead, we rewrite the dictionary in the $u$-coordinate in this paper. 
In the $u$-coordinate,
\begin{subequations}
\begin{align}
A_\mu & \sim \tilcalA_\mu  + \tilA_\mu^{(+)} z + \cdots
= \tilcalA_\mu  + \tilA_\mu^{(+)} \left( \frac{\NGB}{\pi T L}\right)^2 u + \cdots \\
&=:  \calA_\mu + A_\mu^{(+)} u +\cdots~,\\
\Psi &\sim \frac{1}{2} \tilPsi^{(-)} z \ln z +\tilPsi^{(+)} z+ \cdots \\
&=: \half \Psi^{(-)} u \ln u+ \Psi^{(+)} u+ \cdots~, 
\end{align}
\end{subequations}
so that
\begin{subequations}
\label{eq:dict_bdy}
\begin{align}
\bra \calJ^i \ket &= \frac{2}{g^2L} \frac{(\pi T L)^2}{\NGB^3} A_i^{(+)}+(\text{counterterm})~,  \quad (u\text{-coordinate})
\label{eq:dict_current}  \\
\psi &= -\frac{1}{g^2L} \frac{(\pi T L)^2}{\NGB^3} \Psi^{(+)}~, \\
\label{eq:dict_scalar}
J &=  \left( \frac{\pi T L}{\NGB} \right)^2 \Psi^{(-)}~.
\end{align}
\end{subequations}
Note that extra factors of $\NGB$ appear from $T$.

We would like to emphasize that the modification of the AdS/CFT dictionary essentially comes from the nonstandard asymptotic form of the GB black hole. 
It is not because we introduce $\NGB$ in $g_{tt}$. One needs such modifications even if one follows Ref.~\cite{Gregory:2009fj}. The asymptotic behavior in this case is given by 
\begin{subequations}
\begin{align}
ds^2 &\sim \frac{(\pi T' L)^2}{u} \left(-\frac{1}{\NGB^2}dt^2+d\vecx_3^2 \right)+L_\text{eff}^2 \frac{du^2}{4u^2} \\
&=\frac{1}{z}\left(-\frac{1}{\NGB^2}dt^2+d\vecx_3^2 \right)+L_\text{eff}^2\frac{dz^2}{4z^2}~,\\
u&=(\pi T'L)^2z~,
%
\end{align}
\end{subequations}
where $\pi T'L=r_0/L$. The temperature is written as $T'$ to emphasize that $T$ and $T'$ differ due to the difference of $g_{tt}$.
Then, in the $z$-coordinate, one can show
\begin{subequations}
\begin{align}
\bra \calJ^t \ket &= -\frac{2}{g^2L}\tilA_t^{(+)}~,  \quad (z\text{-coordinate})\\
\bra \calJ^i \ket &= \frac{2}{g^2\NGB^2L} \tilA_i^{(+)}+(\text{counterterm})~.
%
\end{align}
\end{subequations}
Note that $\NGB$ and $L$ do not appear in the combination $L_\text{eff}=\NGB L$. 
If one sets $L=g=1$, the charge density reduces to the naive dictionary, but the current density does not.
The AdS/CFT dictionary for a generic scalar field is discussed in \appen{additional}.

\subsection{The order parameter response function (high temperature)}\label{sec:response_high_alpha'}

As discussed in \sect{formal}, we expand 
\begin{subequations}
\begin{align}
A_t &= (\mu_c+\epsmu)(1-u)~, \\
\delta\Psi &= ( F_0+ \la \Fol+\cdots )+
 l^2 ( F_2+\la \Fll+\cdots )+\cdots~. 
%
\end{align}
\end{subequations}
In \sect{critical}, we already obtain $\mu_c, F_0,\Fol$.  
The $O(\la^0)$ solution is obtained in Refs.~\cite{Natsuume:2018yrg,Natsuume:2022kic}:
\begin{subequations}
\begin{align}
F_0 &= -\NGB^3 \delta\psi\, \frac{u}{1+u}~, \\
F_2 & \sim \Cnone\,\frac{q^2-2\epsmu}{8} u\ln u~.
%
\end{align}
\end{subequations}
The remaining solution is
\begin{align}
\Fll & \sim- \frac{1}{16}\Cnone\, \left[ (27+\pi^2-44\ln2)q^2-2(9+\pi^2-12\ln2-12\ln^2 2) \epsmu\right] u\ln u+\cdots~. 
\label{eq:Fll_high}
\end{align}
We are not able to obtain the analytic expression for $\Fll$. However, what we need in the end is the slow falloff at the AdS boundary. The falloff can be evaluated using the method in \appen{LHSC}.
The AdS/CFT dictionary in \sect{dict} is given by
\begin{align}
\delta\Psi \sim \NGB^2\frac{J}{2} u\ln u - \NGB^3\Cone\, u+\cdots~.
%
\end{align}
The slow falloff of $\delta\Psi$ comes from $F_2+\la \Fll$, so
\begin{align}
F_2+\la \Fll \sim  \NGB^2\frac{J}{2} u\ln u~.
%
\end{align}
Then, the source $J$ is given by
\begin{subequations}
\begin{align}
J &= \frac{1}{4}(c_q q^2 - 2c_a\epsmu) \Cnone~, \\
c_q &=1- \half(25+\pi^2-44\ln2)\la~,\\
c_a &=  1 - \half(7+\pi^2-12\ln2-12\ln^2 2)\la~. 
%
\end{align}
\end{subequations}

\subsection{The background}\label{sec:background}

As discussed in \sect{formal}, we expand 
\begin{subequations}
\begin{align}
A_t(u) &= (A_t^{(0)}+\la \Atol+\cdots)+\eps^2 (A_t^{(2)}+\la \Atll+\cdots)+ \cdots~,\\
\Psi(u) &= \eps( \Psi^{(1)}+ \la\Psiol+\cdots )+
 \eps^3( \Psi^{(3)}+\la\Psill+\cdots )+\cdots~. 
%
\end{align}
\end{subequations}
The $O(\la^0)$ solutions are obtained in Refs.~\cite{Herzog:2010vz,Natsuume:2018yrg,Natsuume:2022kic}. The spontaneous condensate solution is given by
\begin{subequations}
\label{eq:back1}
\begin{align}
A_t^{(0)} &= \mu_{0}(1-u)~,\\
\Psi^{(1)} &= -\NGB^3\frac{u}{1+u} \sim -\NGB^3u~,\text{ with } \mu_0=2~, \\
A_t^{(2)} &=  \mu_2(1-u)-\frac{u(1-u)}{4(1+u)}~,\\
\Psi^{(3)} &= \frac{u^2}{12(1+u)^2}-\frac{u\ln(1+u)}{24(1+u)}~, \text{ with }\mu_2=\frac{1}{24}~.
%
\end{align}
\end{subequations}
Note that the Maxwell field introduces an integration constant $\mu_i$ at each order. It is fixed at the next order from the source-free condition of $\Psi^{(i+1)}$.

We need to obtain $O(\la)$ terms. In \sect{critical}, we already obtain
\begin{subequations}
\label{eq:back2}
\begin{align}
\Atol &= \muol (1-u)~,\\
\Psiol &= \Fol \sim O(u^2)~.
%
\end{align}
\end{subequations}
The remaining solutions are
\begin{subequations}
\begin{align}
\Atll &\sim \mull+ \frac{1}{8}(10+\pi^2-8\mull-12\ln2-12\ln^2 2) u+\cdots~, \\
\Psill &\sim \frac{-862-3\pi^2-144\mull+1140\ln2+180\ln^2 2}{576} u\ln u+\cdots~.
%
\end{align}
\end{subequations}
Here, the slow falloff of $\Psill$ is evaluated by the method in \appen{LHSC}.
Then, the source-free condition for $\Psill$ fixes
\begin{align}
\mull = \frac{1}{144}( -862-3\pi^2+1140\ln2+180\ln^2 2 )~.
%
\end{align}
Then, as discussed in \sect{background_formal},
\begin{subequations}
\begin{align}
\epsilon^2 &= 24c_e\epsmu~, \\
c_e &= 1+\frac{1}{6}(862+3\pi^2-1140\ln2-180\ln^2 2)\la~.
%
\end{align}
\end{subequations}
Once $a_0$ is obtained, $b_0$ is given by
\begin{align}
b_0= a_0(\mu_2+\la\mull)~.
%
\end{align}

\subsection{The penetration length}\label{sec:london_alpha'}

For simplicity, we consider $A_y=A_y(x,u)$ with $A_y\propto e^{iqx}$. 
The bulk Maxwell equation becomes
\begin{align}
0=\del_u(\fGB \del_u A_y) -\left( \NGB^2\frac{q^2}{4u} + \frac{|\Psi|^2}{2u^2} \right) A_y~.
%
\end{align}
We impose the boundary conditions (1) regular at the horizon (2) $A_y|_{u=0}=\calA_y$. 
One can rewrite the equation as an integral equation:
\begin{subequations}
\begin{align}
A_y &= \calA_y - \int_0^u \frac{du'}{\fGB(u')} \int_{u'}^1 du''\, V(u'') A_y(u'')~, \\
V &=\NGB^2\frac{q^2}{4u} + \frac{|\Psi|^2}{2u^2}~.
%
\end{align}
\end{subequations}
One can solve the integral equation iteratively. 
At leading order,
\begin{align}
A_y &= \calA_y 
- \calA_y \int_0^u \frac{du'}{\fGB(u')} \int_{u'}^1 du''\, V(u'') +\cdots~,
%
\end{align}
which gives
\begin{align}
\left. 2\del_u A_y \right|_{u=0} 
& =  \frac{-2}{\fGB(0)} \calA_y\int_0^1 du\, V +\cdots~.
%
\end{align}
Then, from the AdS/CFT dictionary \eqref{eq:dict_current}, one obtains
\begin{subequations}
\label{eq:current4}
\begin{align}
\bra \calJ^y \ket &= \left. \frac{2}{\NGB^3}\del_u A_y-\NGB q^2 \calA_y \half(\ln u-2\ln r_0) \right|_{u=0} \\
&= (c_n q^2 - c_s\eps^2) \calA_y~,  \\
c_s & =\half-\frac{1}{4}(25+\pi^2-44\ln2)\la~, \\
c_n &= \NGB\ln r_0= \left(1-\half\la \right)\ln r_0~,
%
\end{align}
\end{subequations}
where  we use the counterterm \eqref{eq:CT}, $\fGB(0)\NGB^2=1$, and the $r_0$-dependence is shown explicitly only for the $\ln r_0$ term.  Also, we use the background solution (\sect{background}):
\begin{align}
|\Psi|^2&= \eps^2 \Psi^{(1)}\left(\Psi^{(1)}+ 2\la\Psiol \right)+\cdots~.
%
\end{align}

%
%
The term $c_s$ represents the supercurrent $\calJ_y^s=-c_s \epsilon^2 \calA_y$. In the GL theory, 
\begin{align}
\calJ_y^s = -2c_0 \epsilon^2 \calA_y~,
%
\end{align}
so  $c_s=2c_0$ as expected. 
The supercurrent increases at finite coupling because  $c_0\epsilon^2$ increases.
The term $c_n$ exists even in the pure Maxwell theory with $\Psi=0$. This term can be interpreted as the normal current or the magnetization current. 
%
%

As the boundary condition at the AdS boundary, we impose the ``holographic semiclassical equation" \cite{Natsuume:2022kic}:
\begin{align}
\del_j \calF^{ij} &=e^2 \bra \calJ^i \ket~.
\label{eq:semiclassical}
\end{align}
Here, all quantities including the $U(1)$ coupling $e$ are the boundary ones. 
In the literature, one often imposes either the Dirichlet or the Neumann boundary conditions. These boundary conditions correspond to $e\to0$ and $e\to\infty$ limits, respectively.
Namely, we impose a ``mixed" boundary condition. The boundary condition gives
\begin{subequations}
\begin{align}
%
q^2 \calA_y &= e^2(c_n q^2-c_s\epsilon^2)\calA_y+e^2\calJ_\text{ext}~, \\
\calA_y &= \frac{e^2}{q^2(1-c_ne^2)+e^2 c_s\epsilon^2} 
\propto \frac{1}{q^2+\mu_m c_s\epsilon^2} =: \frac{1}{q^2+1/\lambda^2}~, \\
\lambda^2 &= \frac{1}{\mu_m c_s\epsilon^2}~,
 \\
\mu_m &= \frac{e^2}{1-c_ne^2}~.
%
\end{align}
\end{subequations}
Then, the net effect of the normal current is to change the magnetic permeability from the vacuum value $\mu_0=e^2$ to $\mu_m$. More explicitly,
\begin{subequations}
\begin{align}
\lambda^2 &= \frac{2}{\mu_m\epsilon^2} \biggl[1+\half(25+\pi^2-44\ln2)\la \biggr] \\
&= \frac{1}{12\mu_m\epsmu}\left[ 1-\frac{1}{6}(787-1008\ln2-180\ln^2 2) \la \right]~, \\
\mu_m &=\frac{e^2}{1-(1-\half\la)e^2\ln r_0}~. 
%
\end{align}
\end{subequations}

\subsection{The order parameter response function (low temperature)}\label{sec:response_alpha'}

We take the gauge $A_u=0$ and perturb around the low-temperature background:
\begin{subequations}
\begin{align}
\Psi &= \bmPsi+\delta\Psi~, \\
A_t &= \bmA_t+a_t~, \\
A_x &= 0 +a_x~,
%
\end{align}
\end{subequations}
where boldface letters indicate the background. We consider the perturbation of the form $e^{iqx}$. 
The $\delta\Psi$ equation is real, so $\delta\Psi^*=\delta\Psi$. In this case, one can set $a_x=0$.
The rest of field equations is given by
\begin{subequations}
\begin{align}
0 &= \del_u^2 a_t - \left[\NGB^2\frac{q^2}{4\fGB u}+\frac{\bmPsi^2}{2\fGB u^2}\right] a_t -\frac{\bmA_t\bmPsi}{\fGB u^2} \delta\Psi~, \\
0 &= \del_u\left(\frac{\fGB}{u}\del_u\delta\Psi\right) + \left[ \frac{\bmA_t^2}{4\fGB u^2}-\NGB^2\frac{q^2}{4u^2}-\frac{m^2}{4u^3} \right]\delta\Psi + \frac{\bmA_t\bmPsi}{2\fGB u^2}a_t~. 
%
\end{align}
\end{subequations}
Set $\eps\to l \eps, q \to l q$. 
Below we consider the case $a_t|_{u=0}=0$ for simplicity (no boundary Maxwell perturbations). 
In this case, one can expand the fields as
\begin{subequations}
\begin{align}
a_t &= 
l (a_t^{(1)}+\la \atll+\cdots)  +\cdots~,\\
\delta\Psi &= (F_0+\la \Fol+\cdots) + l^2(F_2+ \la \Fll+\cdots) +\cdots~.
%
\end{align}
\end{subequations}
We impose the following boundary conditions:
\begin{enumerate}
\item $a_t^{(i)}=0$ at the horizon, no slow falloff except $a_t^{(0)}$, and $a_t|_{u=0}=0$.  
\item $\delta\Psi$: regular at the horizon and the condensate comes only from $F_0$. 
\end{enumerate}
The $O(\la^0)$ solutions are obtained in Refs.~\cite{Herzog:2010vz,Natsuume:2018yrg,Natsuume:2022kic}:
\begin{subequations}
\begin{align}
F_0 &= -\NGB^3\Cnone\,\frac{u}{1+u}~, \\ 
a_t^{(1)} &= - \Cnone\,\eps \frac{u(1-u)}{2(1+u)}~, \\
\frac{F_2}{\Cnone}
&=
 \frac{6q^2+\eps^2}{48}\frac{u\ln u}{1+u} 
- \eps^2 \frac{u\ln (1+u)}{6(1+u)}
+ \eps^2 \frac{u^2}{4(1+u)^2}~. 
%
\end{align}
\end{subequations}
%
%
We need to obtain $O(\la)$ solutions. $\Fol$ is obtained in \sect{critical}.
The remaining solutions are
\begin{subequations}
\begin{align}
\atll 
&\sim \frac{1}{4}\Cnone\, \eps (10+\pi^2-12\ln2-12\ln^2 2) u+\cdots~. \\ 
\Fll &\sim - \frac{1}{16}\Cnone\, \left[ (27+\pi^2-44\ln2)q^2 + \frac{1}{18}(889+6\pi^2-1176\ln2-216\ln^2 2) \eps^2 \right] u\ln u+\cdots~.
%
\end{align}
\end{subequations}
Here, the slow falloff of $\Fll$ is evaluated using the method in \appen{LHSC}.
At $O(q^2)$, the solution $\Fll$ is the same as the high-temperature phase one \eqref{eq:Fll_high}. This is because the field equation at $O(q^2)$ is the same as the high-temperature phase one. This is an expected result from the GL theory, but this is guaranteed by the form of the bulk field equations.

Then, one obtains 
\begin{subequations}
\begin{align}
J &=  \left[ \frac{1}{4}q^2\left\{1-\half(25+\pi^2-44\ln2)\la \right\} 
+\frac{1}{24} \epsilon^2\left\{1-\frac{1}{6}(883+6\pi^2-1176\ln2-216\ln^2 2)\la \right\} \right] \delta\psi \\
&= \left[ \frac{1}{4} q^2\left\{1-\half(25+\pi^2-44\ln2)\la \right\} 
+\epsmu\left\{1-\half(7+\pi^2-12\ln2-12\ln^2 2)\la \right\} \right] \delta\psi \\
&=\frac{1}{4}(c_q q^2+4c_a\epsmu) \delta\psi~,
%
\end{align}
\end{subequations}
as expected from the GL theory.
%
%

\subsection{The conductivity}

We consider the perturbation of the form $A_y \propto e^{-i\omega t+iqx}$ and compute the conductivity. 
The bulk Maxwell equation becomes
\begin{subequations}
\begin{align}
0 &= \frac{1}{\fGB}\del_u(\fGB\del_u A_y) - \frac{1}{\fGB}\left(\NGB^2\frac{q^2}{4u}-\frac{\omega^2}{4\fGB u}+\frac{|\Psi|^2}{2u^2} \right) A_y \\
&= \frac{1}{\fGB}\del_u(\fGB \del_u A_y)-V_0 A_y~.
%
\end{align}
\end{subequations}
 We consider the $q$-dependence in \sect{london_alpha'}, so we set $q=0$ below.
We impose (1) the incoming-wave boundary conditions at the horizon (2) $A_y|_{u=0}=\calA_y$. To incorporate the incoming-wave boundary condition, set the ansatz:
\begin{align}
A_y = g(u)Z(u)~, \quad g(u)=(1-u^2)^{-i\omega/4}~.
%
\end{align}
%
%
Then, the $Z$ equation becomes
\begin{subequations}
\begin{align}
0 &= \frac{1}{F}(FZ')' - V Z~, \\
F &= hg^2~, \\
h &= \fGB~,\\
V &= V_0-\frac{(hg')'}{hg}~.
%
\end{align}
\end{subequations}
One can rewrite the equation as an integral equation:
\begin{align}
Z &=\calA_y -\int_0^u \frac{du'}{F(u')} \int_{u'}^1 du''\, FV(u'')Z(u'')~.
%
\end{align}
One can again solve the equation iteratively. At leading order,
\begin{subequations}
\begin{align}
Z &= \calA_y - \calA_y \int_0^u \frac{du'}{F(u')} \int_{u'}^1 du''\, FV(u'')+\cdots~,\\
2A_y'|_{u=0} &= \left. - \frac{2\calA_y}{F(0)}  \int_0^1 du\, FV+ \cdots \right|_{u=0}~.
%
\end{align}
\end{subequations}
Then, the current is given by
\begin{align}
\bra \calJ^y \ket &= \left. \frac{2}{\NGB^3}\del_u A_y
\right|_{u=0} 
=
\left( \frac{i\omega}{\NGB} - 2c_0 \eps^2\right) \calA_y~.
%
\end{align}
Here, we use 
$\NGB^2F(0)=1$,
and the results in \sect{london_alpha'}.
The conductivity is given by
\begin{subequations}
\begin{align}
\sigma(\omega) = \left. \frac{\bra \calJ^y \ket}{i\omega\calA_y} \right|_{q=0} 
&= \frac{1}{\NGB} + 2\frac{ic_0\epsilon^2}{\omega}+ \cdots \\
&=1+\half\la + \frac{12i}{\omega} \epsmu \left[ 1+\frac{1}{6}(787-1008\ln2-180\ln^2 2)\la \right] \\
&\approx 1+\half\la + \frac{12i}{\omega} \epsmu \left[ 1+0.3043\la \right]~.
%
\end{align}
\end{subequations}
$\Im(\sigma)$ has the $1/\omega$-pole which implies the diverging DC conductivity.
At finite coupling, the supercurrent increases, so the residue of the pole increases as well. 
The finite part of the DC conductivity also increases. 

\subsection{The vortex lattice}

In this subsection, we consider the case where the magnetic field is near the upper critical magnetic field $\Bup$ and consider the vortex lattice.%
\footnote{See, \eg, Refs.~\cite{Natsuume:2022kic,Maeda:2009vf,Albash:2009iq,Montull:2009fe,Keranen:2009re,Domenech:2010nf,Dias:2013bwa} for holographic vortices.}

Previously, we discuss the vortex lattice for a minimal \HSC\ in the SAdS$_5$ background \cite{first}. Here, we use the GB \bh background. The argument is straightforward following Ref.~\cite{first} but is rather involved. However, in Ref.~\cite{first}, we discuss a generic background case and summarize the formulae one needs to evaluate. We use these formulae to obtain main results instead of repeating the exercise. One just needs to replace $f\to \fGB$, needs to insert the factor $\NGB$ appropriately, and evaluate integrals $I_1,I_t,I_L,I_R$ at $O(\la)$. 

The bulk field equations are given in \eq{bulk_eom}. We expand
\begin{subequations}
\begin{align}
\Psi(\vecx,u) &= \eps\Psi^{(1)}+ \eps^3\Psi^{(3)}+\cdots~, \\
A_t(\vecx,u) &= A_t^{(0)}+\eps^2 A_t^{(2)}+\cdots~, \\
A_i(\vecx,u) &= A_i^{(0)}+\eps^2 A_i^{(2)}+\cdots~.
%
\end{align}
\end{subequations}
At zeroth order, 
\begin{align}
A_t^{(0)} = \mu_c(1-u)~, 
A_x^{(0)} = 0~, 
A_y^{(0)} = B_0x~.
%
\end{align}
At first order, one can use separation of variables:
\begin{subequations}
\begin{align}
\Psi^{(1)}(\vecx,u) &= U(u)\psi^{(1)}(x,y)~,\\
U(u)& =\Psi^{(1)}+\la\Psiol~.
%
\end{align}
\end{subequations}
One can solve $U$ and show $B_0=B_{c2}$ as in the high-temperature phase (\sect{high_formal}).

At second order, the Maxwell equation is given by
\begin{subequations}
\begin{align}
0 &= \calL_V A_i^{(2)} - g_i~,\\
\calL_V &= \del_u(\fGB\del_u)-\NGB^2\frac{q^2}{4u}~, \\
g_i &= i\epsilon_i^{~j} q_j \frac{|\Psi^{(1)}|^2}{4u^2}~.
%
\end{align}
\end{subequations}
Obtain 2 independent homogeneous solutions $A_b, A_h$ for $\calL_V A_i^{(2)}=0$ at $O(q^0)$:
\begin{subequations}
\begin{align}
A_h &=1~,\\
A_b &=\half \ln \left(\frac{1-u}{1+u} \right)+ \la u~, \quad \del_uA_b|_{u=0}=-1+\la~,\\
W &:= A_b\del_uA_h-(\del_uA_b)A_h =: \frac{A}{\fGB}~,\\
A &=1~.
%
\end{align}
\end{subequations}
The current is given by
\begin{subequations}
\begin{align}
\bra\calJ_i \ket 
&=\frac{2}{\NGB^3}\del_u A_i^{(2)} +(\text{counterterm}) |_{u=0} \\
&= \calJ_i^{s} +\calJ_i^{n}~.
%
\end{align}
\end{subequations}
Here, the supercurrent $\calJ_i^{s}$ is given by
\begin{subequations}
\begin{align}
\calJ_i^{s} 
&=- i\epsilon_i^{~j} q_j |\psi^{(1)}|^2 \times \frac{1}{\NGB^3} I_1 \\
&=- i\epsilon_i^{~j} q_j c_0 |\psi^{(1)}|^2~,
%
\end{align}
\end{subequations}
where $I_1$ is an integral given by
\begin{align}
\frac{1}{\NGB^3} I_1 &= -\frac{1}{\NGB^3} \frac{\del_u A_b(0)}{A} \int_0^1 du'\,\frac{A_h U^2}{2u'^2} = c_0~.
%
\end{align}
The normal current $\calJ_i^n$ is given by
\begin{subequations}
\begin{align}
\calJ_i^{n} 
&= \frac{1}{\NGB^3\fGB(0)}\NGB^2 q^2  (\ln r_0) \calA_i^{(2)} =: q^2 c_n \calA_i^{(2)}~, \\
c_n &= \NGB\ln r_0~.
%
\end{align}
\end{subequations}
The holographic semiclassical equation \eqref{eq:semiclassical} then gives
\begin{subequations}
\begin{align}
\del_j\calF^{ij} &=e^2\bra\calJ^i\ket~, \\
\to q^2 \calA_i^{(2)} &= e^2q^2 c_n \calA_i^{(2)} + e^2\calJ_i^s~, \\
\to q^2 \calA_i^{(2)} &=  \mu_m \calJ_i^s~, \\
\mu_m &= \frac{e^2}{1-e^2c_n}~.
%
\end{align}
\end{subequations}
$B_2$ is then obtained as
\begin{align}
B_2 = i\epsilon^{ij} q_i \calA_j^{(2)} = -\mu_m c_0 |\psi^{(1)}|^2~.
%
\end{align}
The total $B$ is given by
%
%
\begin{align}
B &=  B_0 +\epsilon^2 B_2 = \Be - \mu_m c_0  |\psi^{(1)}|^2~.
%
\end{align}
This agrees with the analogous expression in the GL theory with the correct coefficient.
The magnetic induction $B$ reduces by the amount $|\psi^{(1)}|^2$, which implies the Meissner effect. 

At  third order, one needs to evaluate the ``orthogonality condition." The orthogonality condition fixes the normalization of the first-order solution $\psi^{(1)}$. One can then evaluate the free energy and determine the vortex lattice configuration. 
The orthogonality condition is given by
\begin{align}
0&= \int d^5x\sqrt{-g}\, J_M^{(2)} A_{(2)}^M~.
%
\end{align}
As discussed in Ref.~\cite{first}, the condition is rewritten as%
\begin{align}
 -2\mu_c^2\bra |\psi^{(1)}|^4 \ket \times I_L
 = \bra B_2 |\psi^{(1)}|^2 \ket \times I_R~, 
%
\end{align}
where
\begin{subequations}
\begin{align}
I_L &= \int_0^1du\, \sqrt{-g} g^{tt}U^2(1-u)I_t~, \\
I_R &= \int_0^1du\, \sqrt{-g} g^{xx}U^2~, \\
I_t &= (1-u)\int_0^{1} du'\, (1-u') g_t(u') -(1-u)\int_0^{u} du'\, g_t(u') - \int_{u}^{1} du'\, (1-u') g_t(u')~,\\
g_t &= \frac{1}{2u^2\fGB}U^2 (1-u)~.
%
\end{align}
\end{subequations}
These integrals can be evaluated as
\begin{subequations}
\begin{align}
I_L &= \frac{1}{384}\left[ \NGB^9-\frac{1}{3}(458+3\pi^2-624\ln2-108\ln^2 2) \la \right]~,\\
I_R 
&= \frac{1}{4} \biggl[ \NGB^5+\half(-20-\pi^2+44\ln2)\la \biggr]~.
%
\end{align}
\end{subequations}
$B_2$ is expressed as
\begin{subequations}
\begin{align}
\B &= \Bup+\B_2 =  \Be-\mu_m c_0 |\psi^{(1)}|^2 \\
\to 
\B_2 &= \Be - \Bup -\mu_m c_0 |\psi^{(1)}|^2~.
%
\end{align}
\end{subequations}
Then, the orthogonality condition becomes
\begin{subequations}
\begin{align}
-2\mu_c^2 \frac{I_L}{I_R} \bra |\psi^{(1)}|^4 \ket
& = (\Be-\Bup) \bra |\psi^{(1)}|^2 \ket -\mu_m c_0 \bra |\psi^{(1)}|^4 \ket~,\\
-2\mu_c^2\frac{I_L}{I_R}  &= -\frac{b_0}{c_0}~. 
\end{align}
\end{subequations}
Thus, the relation reduces to the analogous relation in the GL theory (see,\eg, Appendix~B.1 of Ref.~\cite{first}):
\begin{align}
-\frac{b_0}{c_0}\bra |\psi^{(1)}|^4 \ket
& = (\Be-\Bup) \bra |\psi^{(1)}|^2 \ket - \mu_m c_0 \bra |\psi^{(1)}|^4 \ket~.
%
\end{align}
The rest of the analysis is the same as the GL theory, and the favorable vortex lattice configuration is the triangular lattice even under the $\alpha'$-corrections.

\section{Discussion}

In this paper, we analyze a bulk 5-dimensional minimal \HSC\ and compute various physical quantities at finite coupling. We also identify the dual GL theory exactly. One can understand how various quantities behave in the strong coupling limit from the behaviors of the GL coefficients $\tila, \tilb_0$ (\sect{canonical}). In particular,
\begin{itemize}

\item
$T_c$ takes the highest value in the strong coupling limit.
\item
The GL parameter $\kappa$ increases at finite coupling. Namely, the system becomes a more Type-II \SC\ like material. 
\item
One often says that a finite-coupling correction makes the condensate ``harder," but such a claim must be reexamined. 
This is because previous works typically have 2 problems:
\begin{itemize}
\item
One problem is the naive AdS/CFT dictionary. 
In our example, the condensate actually \textit{increases} at finite coupling (\fig{condensate}). This is not because our system is an exceptional case. If one uses the naive dictionary, the ``condensate" would decrease like common folklore (\appen{ala_soda}). 
We also compute the other quantities using our dictionary and the naive dictionary. 
\textit{The qualitative behaviors of many physical quantities become opposite if one uses our dictionary}, so the use of an appropriate dictionary overwhelms the other effects. 
\item
The other problem is the analysis of the GL potential term only.  
The dual GL theory typically does not have the canonical normalization, and the kinetic term is also corrected. Then, \textit{whether the condensate decreases or not would depend how one normalizes the kinetic term.} In our opinion, the behavior of $\epsilon$ should be compared after one takes the $\la$-independent normalization. 
\end{itemize}
Most previous works do not consider the issues, so one needs to reexamine their results. 


\item
Note that
\begin{itemize}
\item
We analyze only a  \HSC\ with a particular bulk scalar mass, so we do not know if the similar results hold for the other mass.
\item
We analyze only a minimal \HSC, so we do not know if the similar results hold for the other \HSCs.
\item
We analyze a particular higher derivative terms, GB gravity, so we do not know if the similar results hold for the other higher derivative terms. 
\end{itemize}
In order to see if the properties we found are universal or not, it would be interesting to carry out a similar analysis for the other cases either analytically or numerically. 

 Whether the condensate decreases or not depends on the following factors:
\begin{enumerate}
\item
The factor $\NGB$ in the metric
\item
The difference of the AdS/CFT dictionary
\item
The canonical normalization
\end{enumerate}
In our case, these factors together make the condensate increases. But whether the condensate decreases or not depends on the system one chooses (\appen{nonminimal}).

\item
We carry out various computations, but not all computations are independent. For example, we computed the GL kinetic term both in the high temperature phase and in the low temperature phase. But at $O(q^2)$, the bulk field equations are the same, so the results must agree. It would be interesting to study the structure of bulk field equations in more details, and it is important to figure out the number of independent bulk computations. 

\item
The \HSC\ describes a superconductor, but it is different from standard condensed-matter \SCs:
\begin{itemize}
\item
The Cooper pair is formed by the electron-phonon interaction for a \SC, but we couple the complex scalar for the \HSC, and there is no reason to believe that the complex scalar is formed from fermions. (However, there are a few attempts to study the Cooper pair formation in AdS/CFT \cite{Hartman:2010fk}.) 
In addition, $\la$ has no relation to the electron-phonon coupling, and the meaning of $\la$ remains unclear in condensed-matter systems. 
\item
In AdS/CFT, the gravitational theory is dual to a non-Abelian plasma such as the quark-gluon plasma. Such a plasma plays the role of the medium, and the plasma is strongly-coupled. Here, the meaning of $\la$ is clear, and the effect of strongly-coupled medium is seen from finite-coupling corrections, \eg, the magnetic permeability. It changes from the vacuum value $\mu_0=e^2$ due to the medium effect as we have seen.
\item
The above arguments do not imply that the \HSC\ is useless. One can still learn interesting lessons from the \HSC. Also, the \HSC\ is a new kind of \SC\ which can occur in a non-Abelian plasma.
\end{itemize}

\item
Finally, we take the probe limit. It is interesting to take the backreaction into account to see how our results change.  
\end{itemize}

\acknowledgments

I would like to thank Takashi Okamura for his continuous suggestions and interest throughout the work.
This research was supported in part by a Grant-in-Aid for Scientific Research (17K05427) from the Ministry of Education, Culture, Sports, Science and Technology, Japan. 

\appendix

\small

\section{Supplementary information}

\subsection{Physical quantities by the naive dictionary}\label{sec:ala_soda}

We cannot compare our results with the results in Ref.~\cite{Gregory:2009fj} directly because our bulk scalar mass is different from theirs. Instead, we give the results if one follows the reference. Namely, we use the metric
\begin{subequations}
\label{eq:metric_soda}
\begin{align}
ds^2 &= \left(\frac{r_0}{L}\right)^2\frac{1}{u}(-\fGB dt^2+d\vecx_3^2)+L^2\frac{du^2}{4u^2\fGB}~, \\
 &= \frac{1}{z}(-\fGB dt^2+d\vecx_3^2)+L^2\frac{dz^2}{4z^2\fGB}~, 
%
\end{align}
\end{subequations}
where $u=(r_0/L)^2 z$, and we use the naive dictionary:
\begin{subequations}
\begin{align}
\Psi &\sim \frac{\tilde{J}}{2} z\ln z-\tilde{\psi} z~, \\
A_i &\sim \tilcalA_i + \frac{\tilde{J}^i}{2} z~.
%
\end{align}
\end{subequations}
For the metric, the temperature is $\pi T' L=r_0/L$, and we fix $\pi T'=1,L=1$ below. 
%
%

Then, we get
%
%
\begin{subequations}
\begin{align}
c_0 &=\frac{1}{4} \biggl[ 1-\half(22+\pi^2-44\ln2)\la \biggr]\\
&\approx \frac{1}{4}(1-0.6856\la)~, \decrease  
\\
a_0 &= \half \biggl[ 1-\half(6+\pi^2-12\ln2-12\ln^2 2)\la \biggr]\\
&\approx \half(1-0.8932\la)~, \decrease  
\\
b_0 &=\frac{1}{48} \biggl[1-\frac{1}{6}(862+6\pi^2-1176\ln2-216\ln^2 2)\la \biggr]\\
&\approx \frac{1}{48}(1-0.3831\la)~, \decrease 
\\
\mu_c &=2+(10-12\ln2)\la \approx 2+1.682\la~,\\
\mu_m &=\frac{e^2}{1-c_n e^2}~,  \\
c_n&=\left(1-\la\right)\ln r_0~.
%
\end{align}
\end{subequations}
Thus,
\begin{subequations}
\begin{align}
\epsilon^2 &= 24\epsmu \biggl[ 1-\frac{1}{6}(-844-3\pi^2+1140\ln2+180\ln^2 2) \la \biggr] \\
&\approx 24\epsmu(1-0.5101\la)~, \decrease
\\
\xi_>^2 
&= \frac{1}{-2\epsmu}\biggl[ 1+( -8+16\ln2-6\ln^2 2) \la \biggr] \\ 
&\approx \frac{1}{-2\epsmu}(1+0.2076\la)~, \increase 
\\
\lambda^2 
&= \frac{1}{12\mu_m\epsmu}\biggl[ 1+\frac{1}{6}(-778+1008\ln2+180\ln^2 2)\la \biggr]\\
&\approx \frac{1}{12\mu_m\epsmu}(1+1.196\la)~, \increase 
\\
\kappa^2 
&= \frac{1}{6\mu_m}\biggl[ 1+\frac{1}{6}( -730+912\ln2+216\ln^2 2 )\la \biggr] \\
&\approx \frac{1}{6\mu_m}(1+0.9880\la)~, \increase
\\
\Bup 
&= 2\epsmu \biggl[1-( -8+16\ln2-6\ln^2 2) \la \biggr] \\
&\approx 2\epsmu(1-0.2076\la)~.   \decrease
\\
\Bc^2 
&= 12\mu_m \epsmu^2 \biggl[1-\frac{1}{6}( -826+1104\ln2+144\ln^2 2 )\la \biggr] \\
&\approx 12\mu_m\epsmu^2 (1-1.403\la)~. \decrease 
%
\end{align}
\end{subequations}
At finite coupling, the ``condensate" $\epsilon$ decreases, namely, $\la$ makes the ``condensate" harder as is often stated. 
Also, the qualitative behaviors of $\xi, \lambda,B_{c2}, B_c$ are opposite from the results in the text.  

In the canonical form,
\begin{subequations}
\begin{align}
f &= |D_i\phi|^2 -\tila|\phi|^2+\frac{\tilb}{2}|\phi|^4+\cdots~, \\
\tila &= \frac{a}{c_0} 
= 2\epsmu \left[ 1- (-8+16\ln2-6\ln^2 2)  \la  \right]  \\
&\approx 2\epsmu(1-0.2076\la)~, \decrease 
\\
\tilb_0 &= \frac{b_0}{c_0^2} 
= \frac{1}{3}\left[ 1+ \frac{1}{6}(-730+912\ln2+216\ln^2 2) \la  \right] \\
&\approx \frac{1}{3}(1+0.9880\la)~. \increase 
%
\end{align}
\end{subequations}
The condensate changes as
\begin{subequations}
\begin{align}
|\phi|^2 &=\frac{\tila}{\tilb_0} 
= 6 \epsmu \left[ 1- \frac{1}{6}(-778+1008\ln2+180\ln^2 2) \la \right] \\
&\approx 6\epsmu(1-1.196\la)~.
\decrease 
%
\end{align}
\end{subequations}
In the canonical form, $\la$ makes the ``condensate" harder again.

\subsection{The additional contributions from our dictionary}\label{sec:additional}

One can easily estimate how our dictionary changes the naive results. Consider a generic bulk scalar field with the asymptotic behavior 
\begin{align}
\Psi &\sim \tilPsi^{(-)} z^{\Deltam/2} + \tilPsi^{(+)} z^{\Deltap/2}~.
%
\end{align}
In this case, the dictionary is given by
\begin{subequations}
\begin{align}
\psi &= \frac{\Deltap-\Deltam}{\NGB L}\tilPsi^{(+)}~,  \quad (z\text{-coordinate})\\
J &= \tilPsi^{(-)}~.
\end{align}
\end{subequations}
\begin{subequations}
In the radial coordinate $u=(r_0/L)^2z$,
\begin{align}
\Psi &\sim \tilPsi^{(-)} \left(\frac{r_0}{L} \right)^{-\Deltam} u^{\Deltam/2} + \tilPsi^{(+)} \left(\frac{r_0}{L} \right)^{-\Deltap}u^{\Deltap/2} \\
 &\sim \Psi^{(-)} u^{\Deltam/2} + \Psi^{(+)} u^{\Deltap/2}~.
%
\end{align}
\end{subequations}
Let us rewrite the dictionary in the $u$-coordinate. Below we ignore numerical factors.
\begin{enumerate}
\item
In the $u$-coordinate, one obtains
\begin{subequations}
\label{eq:dict_condensate}
\begin{align}
%
\psi &\propto\frac{1}{\NGB L}\left(\frac{r_0}{L} \right)^{\Deltap}\Psi^{(+)} 
\quad  (u\text{-coordinate}) \\
&=\frac{(\pi T L)^{\Deltap}}{\NGB^{\Deltap+1} L} \Psi^{(+)} = \frac{1}{\NGB^{\Deltap+1}} \Psi^{(+)}~. 
%
\end{align}
\end{subequations}
Here, $\pi T=\NGB r_0/L$, and we set $\pi T=L=1$ in the last expression.
Note that the extra factor of $\NGB^{\Deltap}$ appears from $T$.
Similarly, 
\begin{align}
J &=  \left( \frac{\pi T L}{\NGB} \right)^{\Deltam} \Psi^{(-)}~.
\label{eq:dict_source}
\end{align}

\item
On the other hand, if one uses the metric \eqref{eq:metric_soda} of Ref.~\cite{Gregory:2009fj} and uses the naive dictionary,
\begin{subequations}
\label{eq:naive_scalar}
\begin{align}
\psi &\propto\frac{1}{L}\left(\frac{r_0}{L} \right)^{\Deltap}\Psi^{(+)} = \Psi^{(+)}~,
\quad  (u\text{-coordinate}) \\
J &=  \left( \frac{r_0}{L} \right)^{\Deltam} \Psi^{(-)} = \Psi^{(-)}~.
%
\end{align}
\end{subequations}
Here, $\pi T'=r_0/L$, and we set $\pi T'=L=1$ in the last expressions.
\end{enumerate}

%
%

Below we count only the factor $\NGB$ and ignore numerical factors and dimensionful quantities. 
Let us use the metric \eqref{eq:metric_soda}, and consider the high-temperature phase. Suppose that in the $u$-coordinate one gets 
\begin{align}
\Psi^{(-)} = (c_q q^2-c_a\epsmu)\Psi^{(+)}~.
%
\end{align}
Using the naive dictionary \eqref{eq:naive_scalar}, one would interpret the result as $J=(c_q q^2-c_a \epsmu)\delta\psi$. But if one uses our metric \eqref{eq:GB_u} and our dictionary \eqref{eq:dict_condensate} and \eqref{eq:dict_source}, the result is interpreted as
\begin{subequations}
\begin{align}
\NGB^{\Deltam} J &\sim \NGB^{\Deltap+1}(\NGB^2 c_qq^2-c_a\epsmu)\delta\psi~,\\
\to J &\sim \NGB^{2\Deltap-3}(\NGB^2 c_qq^2-c_a\epsmu)\delta\psi~,
%
\end{align}
\end{subequations}
where $\Deltap+\Deltam=4$ is used. We also replace $q\to \NGB q$. This factor comes from the fact that our $g_{tt}$ differs from Ref.~\cite{Gregory:2009fj} by the factor $\NGB^2$.
Then, 
\begin{align}
c_0\sim \NGB^{2\Deltap-1}c_q~, a_0 \sim \NGB^{2\Deltap-3}c_a~.
%
\end{align}

In the low-temperature phase, we normalize
\begin{align}
\Psi^{(1)} \sim -u~.
%
\end{align}
Then, 
\begin{subequations}
\begin{align}
A_t(u) &= (A_t^{(0)}+\cdots)+(\NGB^{\Deltap+1}\eps)^2 (A_t^{(2)}+\cdots)+ \cdots~,\\
\Psi(u) &= (\NGB^{\Deltap+1}\eps) ( \Psi^{(1)}+\cdots )+\cdots~.
%
\end{align}
\end{subequations}
so that
\begin{subequations}
\begin{align}
\mu&=\mu_c + (\NGB^{\Deltap+1}\eps)^2 \\
\epsilon^2 &=\NGB^{-2\Deltap-2} c_e\epsmu = \frac{a_0}{b_0}\epsmu~,\\
b_0 &= \NGB^{4\Deltap-1} \frac{c_a}{c_e}~.
%
\end{align}

\end{subequations}

Thus, the free energy and physical quantities behave as
\begin{subequations}
\label{eq:N_dict}
\begin{align}
f &= \NGB^{2\Deltap-3} \{ \NGB^2 c_0|D\psi|^2 - a_0\epsmu |\psi|^2 \} + \half \NGB^{4\Deltap-1} b_0 |\psi|^4+\cdots~, \\
\eps^2 &\sim \NGB^{-2\Deltap-2}~,\\
\xi^2 &\sim \NGB^2~, \quad
\lambda^2 \sim \NGB^3~, \quad
\kappa^2 \sim \NGB~.
%
\end{align}
\end{subequations}
Note that we only estimate the powers of $\NGB$ which comes from our dictionary. The coefficients $c_0,a_0,b_0$ have $O(\la)$ contributions as well [see \eq{naive_result}].

\begin{enumerate}
\item
In our case, $\Deltap=\Deltam=2$, so
\begin{subequations}
\begin{align}
f &= \NGB^3|D\psi|^2 - \NGB |\psi|^2 + \NGB^7 |\psi|^4+\cdots~, \\
\eps^2 &\sim \NGB^{-6} \sim 1+3\la~,\\
\xi^2 &\sim \NGB^2~, \quad
\lambda^2 \sim \NGB^3~, \quad
\kappa^2 \sim \NGB~.
%
\end{align}
\end{subequations}
The naive results are
\begin{subequations}
\label{eq:naive_result}
\begin{align}
\eps^2 &\approx (1-0.5\la) 24\epsmu~,\\
\xi_>^2 &\approx(1+0.2\la)  \frac{1}{-2\epsmu}~,\\
\lambda^2 &\approx(1+1.2\la) \frac{1}{12\mu_m\epsmu}~,
%
\end{align}
\end{subequations}
(see \appen{ala_soda}). In particular, $\epsilon$ decreases at finite coupling, or $\la$ makes the condensate ``harder" as is often stated.  
But adding the contributions from our dictionary \eqref{eq:N_dict} to the naive results gives
\begin{subequations}
\begin{align}
\eps^2 &\approx (1+3\la)(1-0.5\la) 24\epsmu \propto 1+2.5\la~,\\
\xi_>^2 &\approx(1-\la)(1+0.2\la)  \frac{1}{-2\epsmu} \propto 1-0.8\la~,\\
\lambda^2 &\approx(1-1.5\la)(1+1.2\la) \frac{1}{12\mu_m\epsmu}  \propto 1-0.3\la~.
%
\end{align}
\end{subequations}
Namely, the contributions from our dictionary are relatively large so that 
the qualitative behaviors of these physical quantities become opposite from the naive results.

\item
As another example, Ref.~\cite{Gregory:2009fj} takes $m^2=-3$, or $(\Deltap,\Deltam)=(3,1)$, so
\begin{subequations}
\begin{align}
f &= \NGB^5|D\psi|^2 - \NGB^3 |\psi|^2 + \NGB^{11} |\psi|^4+\cdots~, \\
\eps^2 &\sim \NGB^{-8} \sim 1+4\la~.
%
\end{align}
\end{subequations}

\item
Finally, in the canonical form, 
\begin{subequations}
\begin{align}
|\phi|^2 &:= \NGB^{2\Deltap-1}|\psi|^2~, \\
f &= |D\phi|^2-\frac{1}{\NGB^2}|\phi|^2 +\NGB|\phi|^4+\cdots~,\\
\epsilon^2 &\sim \NGB^{-3}~.
%
\end{align}
\end{subequations}
\end{enumerate}

\subsection{Extracting falloffs}\label{sec:LHSC}

Following the procedure in \sect{bulk_alpha'}, one can obtain bulk results. However, in some cases, one may not be able to obtain analytic solutions. However, what one would like in the end are the falloffs at the AdS boundary. The slow falloffs have simple expressions \cite{Natsuume:2018yrg}.

We solve the following differential equation:
\begin{subequations}
\begin{align}
\calL\varphi&=\nj~,
\label{eq:inhomo} \\
\calL &=\del_u(p(u)\del_u)~.
%
\end{align}
\end{subequations}
Denote two independent solutions of the homogeneous equation $\calL\varphi=0$ as $\varphi_1$ and $\varphi_2$. We assume that $\varphi_1$ satisfies the boundary condition at the horizon $u=1$. The solution of the inhomogeneous equation \eqref{eq:inhomo} which is regular at the horizon is given by
\begin{align}
\varphi (u) &=
 - \varphi_1(u) \int^u_0 du'\, \frac{\nj(u') \varphi_2(u')}{p(u')W(u')}
 - \varphi_2(u) \int^1_u du'\, \frac{\nj(u') \varphi_1(u')}{p(u')W(u')}~,
\label{eq:inhomo_sol} 
%
\end{align}
where $W$ is the Wronskian
$W(u) := \varphi_1 \varphi_2' - \varphi_1' \varphi_2 $.

For example, for the $\delta\Psi$-perturbation, 
\begin{subequations}
\begin{align}
%
\varphi_1 &= \frac{u}{1+u}~, \\
\varphi_2 &= \frac{u}{1+u} \ln\left[ \frac{u}{(1-u)^2}\right] \sim u \ln u~, \\
p(u) &= \frac{f}{u}~,\quad
pW =1~.
%
\end{align}
\end{subequations}

Even if the integral \eqref{eq:inhomo_sol} is difficult to evaluate or has a cumbersome expression, one can  extract a falloff. Suppose that $\varphi_2$ has the appropriate falloff. Then, near the AdS boundary $u\to\delta$, 
\be
\varphi(\delta) \sim - \varphi_2(\delta) \int^1_{\delta} du\, \nj(u)\varphi_1(u)~.
\label{eq:inhomo_bdy}
\ee
This integral essentially gives the falloff coefficients we want. 

The $\delta$-dependence in the integral essentially has no contribution from the following reason. First, the integral may or may not converge:
\begin{enumerate}
\item
When it converges, one can take the $\delta\to0$ limit since the $\delta$-dependence in the integral does not produce an appropriate falloff when it is combined with $\varphi_2(\delta)$; it gives a subleading falloff.
\item
When it diverges, simply discard the $\delta$-dependence in the integral since again it does not produce an appropriate falloff.%
\footnote{There may be an exception. The $\delta$-dependence in the integral may produce an appropriate falloff when it is combined with the \textit{subleading} term of $\varphi_2(\delta)$.} Even if it diverges as $\delta\to0$, the expression \eqref{eq:inhomo_sol} itself does not.
\end{enumerate}

For  example, consider the $\delta\Psi$-perturbation at high temperature. The slow falloff of  $F_2$ is given by
\begin{subequations}
\begin{align}
\nj &= \biggl[ \frac{\epsmu(1-u)^2}{fu^2}-\frac{q^2}{4u^2} \biggr] F_0~, \\
\frac{J}{2} &= -\int^1_0 du\, \nj \frac{u}{1+u} \\
&=C_1\frac{q^2-2\epsmu}{8}~.
%
\end{align}
\end{subequations}


\section{Nonminimal \HSCs\ at finite coupling}\label{sec:nonminimal}

As we saw in the text,%
\footnote{This appendix is added after publication. }
there are several places where $\NGB$ appears, but they were mostly ignored in previous works:
\begin{enumerate}
\item
Previous works did not consider the Minkowski metric as the boundary metric so that the metric has a factor $\NGB.$
\item
Previous works did not use the appropriate AdS/CFT dictionary. This gives additional factors of $\NGB$.
\item
Previous works did not take the canonical normalization for the dual GL theory. This also gives additional factors of $\NGB$.
\end{enumerate}
For our system in the text, these factors add up together and the condensate increases under higher-derivative corrections.
But probably this is not always the case.
As an example, we consider ``nonminimal" \HSCs\ in this appendix and check the behavior of the condensate.

Ref.~\cite{Herzog:2010vz} studies the analytic solution for a class of nonminimal holographic superconductors (St\"{u}ckelberg holographic superconductor):
\begin{subequations}
\begin{align}
S_\text{m} &= -\frac{1}{g^2} \int d^5x \sqrt{-g} \biggl\{ \frac{1}{4}F_{MN}^2 + K|D_M\Psi|^2+V \biggr\}~,\\
K &= 1+A|\Psi|^2~,\quad
V= m^2 |\Psi|^2+B |\Psi|^4~.
%
\end{align}
\end{subequations}
$A$ and $B$ are bulk parameters. The original \HSC\ is the $A=B=0$ case which we call the ``minimal" \HSC.
The arbitrary values may not be allowed for $A$ and $B$. If $A<0$, $\Psi$ may become a (unitary) ghost. If $B<0$, the potential may not be bounded below. For simplicity, we set $A,B>0$. 

The dual GL theory in the strong coupling limit has been identified in Ref.~\cite{first}:
\begin{subequations}
\begin{align}
f &= c_0 |D_i\psi|^2-a_0\epsmu|\psi|^2+\frac{b_0}{2}|\psi|^4+\cdots~, \\
&= \frac{1}{4}|D_i\psi|^2-\frac{\epsilon_\mu}{2}|\psi|^2+\frac{1+4A+4B}{96}|\psi|^4+\cdots~.
%
\end{align}
\end{subequations}
The nonminimal \HSCs\ have a larger GL parameter $\kappa$ than the minimal case (when $A,B>0$) \cite{first}:
\begin{align}
\kappa^2 &=\frac{1+4A+4B}{6 \mu_m }~.
%
\end{align}

We compute the finite-coupling corrections for the nonminimal \HSCs.
The finite-coupling correction also makes $\kappa$ larger.  So, one expects that the nonminial \HSCs\ at finite coupling have a larger $\kappa$.

Even for the nonminimal \HSCs, the critical point is the same as the minimal \HSC:
\begin{align}
\mu_c=2+(10-12\ln2)\la \approx 2+1.682\la~.
%
\end{align}
Also, the GL coefficients $a_0,c_0$ are the same as well \cite{first}:
\begin{subequations}
\begin{align}
c_0 &=\frac{1}{4} \biggl[ 1-\half(25+\pi^2-44\ln2)\la \biggr]~,\\
a_0 &= \half \biggl[ 1-\half(7+\pi^2-12\ln2-12\ln^2 2)\la \biggr]~.
%
\end{align}
\end{subequations}
This is because these quantities can be derived from the linear perturbation problem in the high-temperature phase. The nonlinear terms $A,B$ do not affect the linear perturbation problem in the high-temperature phase, so the perturbation equation has the same form as the minimal case.
Then, one only needs to obtain $b_0$. For that purpose, it is enough to compute the low-temperature background. 

Again, we construct the background as a double-series expansion both in $\epsilon$ and $\la$:
\begin{subequations}
\begin{align}
A_t(u) &= (A_t^{(0)}+\la \Atol+\cdots)+\eps^2 (A_t^{(2)}+\la \Atll+\cdots)+ \cdots~,\\
\Psi(u) &= \eps( \Psi^{(1)}+ \la\Psiol+\cdots )+
 \eps^3( \Psi^{(3)}+\la\Psill+\cdots )+\cdots~. 
%
\end{align}
\end{subequations}
The Maxwell solutions remain the same as the minimal one:
\begin{subequations}
\begin{align}
A_t^{(2)} &\sim  \mu_2 - \frac{1}{4}(4+\mu_2) u+\cdots~,\\
\Atll &\sim \mull+ \frac{1}{8}(10+\pi^2-8\mull-12\ln2-12\ln^2 2) u+\cdots~.
%
\end{align}
\end{subequations}
The differences are the values of $\mu_2,\mull$.

For the bulk scalar field, $\Psi^{(1)}$ and $\Psiol$ remain the same. But the source-free condition for $\Psi^{(3)}$ fixes 
\begin{align}
\mu_2 =\frac{1+4A+4B}{24}~.
%
\end{align}
Also, the source-free condition for $\Psill$ fixes
\begin{align}
%
\mull &= -\frac{1}{144}
   \left( 862+3 \pi ^2-1140  \ln 2 -180 \ln ^22 \right)+
\nonumber \\
   &-\frac{1}{36} A \left(-476+3\pi^2+480\ln2+252\ln^2 2 \right)
   -\frac{1}{36} B \left( 208+3\pi^2-348\ln2+36\ln^2 2 \right)~.
%
\end{align}
As discussed in the text, the GL coefficient $b_0$ is then given by
\begin{subequations}
\begin{align}
b_0 =& a_0(\mu_2+\la\mull) 
\nonumber \\
=& \frac{1}{48} (1+4A+4B)
-\frac{1}{288} \la\left(883+6 \pi ^2-1176 \ln 2-216 \ln ^22 \right)
 \nonumber \\
& -\frac{1}{72} \la A \left(-455+6 \pi ^2+444 \ln 2+216 \ln ^22 \right)
 -\frac{1}{72} \la B  \left(229+6 \pi ^2-384 \ln 2\right) \\
\approx& \frac{1}{48} (1+4A+4B)-\la (0.08090+0.2188 A +0.3062 B)~.
%
\end{align}
\end{subequations}
Then, one can determine physical quantities. 

The correlation length $\xi$ remains the same as the minimal case. This is because $a_0, c_0$ are the same as the minimal case. Namely, the correlation length increases at strong coupling, and the correlation between longer distance is possible at strong coupling.  

The GL parameter is given by
\begin{subequations}
\begin{align}
\kappa^2 =& \frac{1}{2\mu_m}\frac{b_0}{c_0^2} \\
=& \frac{1+4A+4B}{6 \mu_m }
+\la\frac{-733+912 \ln 2 +216 \ln ^22}{36 \mu_m }
\nonumber \\
&+\la\frac{A \left(605-708 \ln 2-216 \ln ^22\right)+B (-79+120 \ln 2)}{9 \mu_m } \\
\approx& \frac{1+4A+4B}{6 \mu_m }+\frac{\la}{\mu_m}(0.08134+ 1.164 A+ 0.4642 B)~.
%
\end{align}
\end{subequations}
As expected, $\kappa$ increases at a finite $\la>0$ when $A,B>0$.

The ``condensate" is given by 
\begin{subequations}
\begin{align}
\epsilon^2 =\frac{a}{b_0} 
=& \frac{1}{\mu_2+\la \mull}\epsmu \\
=& \frac{24\epsmu}{1+4A+4B}
+4\la \frac{862+3 \pi ^2-1140  \ln 2 -180 \ln ^22
}{(1+4A+4B)^2}\epsmu \nonumber\\
 & +16\la \frac{A (-476+3\pi^2+480\ln2+252\ln^2 2)
+ B (208+3\pi^2-348\ln2+36\ln^2 2)
 }{(1+4A+4B)^2} \epsmu \\
\approx& \frac{24 \epsilon_\mu }{1+4A+4B}
+4\la \frac{14.94+29.57A+54.76B}{(1+4A+4B)^2}\epsilon_\mu~.
%
\end{align}
\end{subequations}
In the $A,B\to 0$ limit, these results reduce to the ones for the minimal \HSC.
In the $\la\to0$ limit, these results reduce to the strong coupling ones in Ref.~\cite{first}.

The ``condensate" increases at a finite $\la>0$ when $A,B>0$. 
But one should consider the condensate in the canonical normalization.
In the canonical normalization, 
%
%
the condensate is given by
\begin{subequations}
\begin{align}
\tilde{\epsilon}^2 =\frac{a}{b_0}c_0 =&
\frac{6}{1+4A+4B}\epsmu
+ \la \frac{ 787-1008\ln2-180 \ln ^22 }{(1+4A+4B)^2}\epsmu
\nonumber \\
&+\la \frac{-A \left(2204-2448 \ln 2-1008 \ln ^22 \right)
+B \left(532-864 \ln 2 +144 \ln ^22 \right)
   }{(1+4A+4B)^2}\epsmu\\
\approx& \frac{6}{1+4A+4B}\epsmu+ \la\frac{1.826-22.88A+2.306 B}{(1+4A+4B)^2}\epsmu~. 
%
\end{align}
\end{subequations}
The condensate increases for $B>0$ as in the minimal \HSC. But it can decrease by choosing $A$ appropriately. 
Thus, the condensate can increase or decrease depending on the system, and there is no universality.

\end{document}